\documentclass[article, shortnames, nojss]{jss} 

\usepackage{orcidlink,thumbpdf,lmodern}

\usepackage{framed}

\usepackage{caption}
\usepackage{enumitem}
\usepackage{array, booktabs} 
\usepackage{multicol} 
\usepackage{subcaption, graphicx}
\usepackage{amsmath}
\usepackage{bm}
\usepackage{longtable}
\usepackage{amssymb}

\makeatletter
\renewcommand{\paragraph}{%
  \@startsection{paragraph}{4}%
  {\z@}{1ex \@plus 1ex \@minus .2ex}{-1em}%
  {\normalfont\normalsize\bfseries}%
}
\makeatother

\usepackage{pifont}
\newcommand{\cmark}{\ding{51}}
\newcommand{\xmark}{--}

\author{Daniele Girolimetto~\orcidlink{0000-0001-9387-1232}\\University of Padova
   \And Jeroen Rombouts~\orcidlink{0000-0003-2255-4875}\\ESSEC Business School \AND
   Ines Wilms~\orcidlink{0000-0003-3269-4601}\\Maastricht University \And
   Yangzhuoran Fin Yang~\orcidlink{0000-0002-1232-8017}\\Maastricht University}
\Plainauthor{Daniele Girolimetto, Jeroen Rombouts, Ines Wilms, Yangzhuoran Fin Yang}

\title{\pkg{FoReco} and \pkg{FoRecoML}: A Unified  Toolbox for Forecast Reconciliation in \proglang{R}}
\Plaintitle{FoReco and FoRecoML: A Unified  Toolbox for Forecast Reconciliation in R}
\Shorttitle{\pkg{FoReco} and \pkg{FoRecoML}: Unified Forecast Reconciliation in \proglang{R}}

\Abstract{
 Forecast reconciliation has become key to improving the accuracy and coherence of forecasts for linearly constrained multiple time series,  such as hierarchical and grouped series. 
 Yet, comprehensive software  that jointly covers cross-sectional, temporal, and cross-temporal reconciliation has so far been lacking. 
 The \proglang{R} packages \pkg{FoReco} and \pkg{FoRecoML} address this gap by offering a comprehensive and unified framework. 
 The packages respectively implement classical and regression-based linear reconciliation approaches, and  non-linear approaches based on machine learning  for cross-sectional, temporal and cross-temporal frameworks. 
Designed for accessibility and flexibility, these packages provide sensible default options that allow new users to apply reconciliation methods with minimal effort, while still giving expert users full control to explore state-of-the-art extensions through customized settings. With this dual focus, \pkg{FoReco} and \pkg{FoRecoML} are versatile tools for practitioners and researchers  working on forecast reconciliation.
}

\Plainkeywords{Aggregation, Coherence, Forecast reconciliation,  Linearly constrained multiple time series, Machine learning}

\Address{
  Daniele Girolimetto\\
  Department of Statistical Sciences\\
  University of Padova\\
  Via Cesare Battisti~241\\
  35121 Padova, Italy\\
  E-mail: \email{daniele.girolimetto@unipd.it}\\
  URL: \url{https://danigiro.github.io/}\\ \\
  Jeroen Rombouts\\
  Department of Information Systems, Data Analytics and Operations\\
  ESSEC Business School\\
   Av. Bernard Hirsch~3\\
  95000 Cergy, France\\
  E-mail: \email{rombouts@essec.edu}\\
  URL: \url{https://www.jeroenvkrombouts.com/}\\ \\
  Ines Wilms\\
  Department of Quantitative Economics, School of Business and Economics\\
  Maastricht University\\
  Tongersestraat 53\\
  6211 LM Maastricht, the Netherlands\\
  E-mail: \email{i.wilms@maastrichtuniversity.nl}\\
  URL: \url{https://sites.google.com/view/iwilms}\\ \\
  Yangzhuoran Fin Yang\\
  Department of Data Analytics and Digitalisation, School of Business and Economics\\
  Maastricht University\\
  Tongersestraat 53\\
  6211 LM Maastricht, the Netherlands\\
  E-mail: \email{yangzhuoran.yang@maastrichtuniversity.nl}\\
  URL: \url{https://yangzhuoranyang.com/}\\[-1em]
}

\begin{document}

\section[Introduction]{Introduction} \label{sec:intro}

In this paper, we introduce the forecast reconciliation packages \pkg{FoReco} and \pkg{FoRecoML} for \proglang{R} \citep{R}. 
Forecast reconciliation adjusts forecasts for linearly constrained multiple time series (such as hierarchical or grouped series, or series observed at different temporal frequencies) so that they are coherent with respect to the underlying constraints, improving both accuracy and consistency for informed decision making.
The contributions of the packages are threefold.
First, \pkg{FoReco} and \pkg{FoRecoML}  are the first to offer functionality for forecast reconciliation methods across cross-sectional, temporal and cross-temporal frameworks. 
Second, the packages provide a comprehensive set of forecast reconciliation approaches, including classical (e.g., top-down,  bottom-up and middle-out) and regression based reconciliation methods  -- in \pkg{FoReco} -- as well as non-linear reconciliation methods using machine learning -- in \pkg{FoRecoML}.
A third key contribution is their unified design, which enables easy-to-use forecast reconciliation functions built on the same philosophy, regardless of the reconciliation framework or method.

Many forecasting tasks involve multiple time series that must satisfy linear constraints. Forecast reconciliation is crucial in this context, as it delivers coherent forecasts across all the system, thereby supporting aligned decision making.   
For \textit{cross-sectional} frameworks (e.g., hierarchical or grouped structures), reconciliation helps correct individual forecasts to be coherent; think of aggregating regional-level to national tourism demand.
For \textit{temporal} frameworks, a single series can be observed at different temporal frequencies; think of aggregating quarterly tourism demand semi-annually and anually.
For \textit{cross-temporal} frameworks, the aggregation spans across both cross-sectional and temporal constraints.
Readers new to forecast reconciliation may consult Chapter 11 in \cite{hyndman2021fpp3} for a classical textbook introduction or \cite{athanasopoulos2023review} for a recent review.

Cross-sectional, temporal and cross-temporal reconciliation are wide-spread in real-world forecasting problems, highlighting the need for software for all frameworks. 
Table \ref{tab:pkg-overview} provides an overview of existing \proglang{R} and \proglang{Python} \citep{Python} packages alongside their key forecast reconciliation functionalities. 
Support for the three forecast reconciliation frameworks is scattered across  packages.
Most packages only adopt 
one reconciliation framework 
(cross-sectional, e.g., \citealp{Hyndman2011-fh}: \pkg{hts}, \pkg{ProbReco}, \pkg{gtop}, \pkg{scikit-hts}, \pkg{reconcile}, \pkg{Fable}, \pkg{fabletools}, \pkg{pyhts}, 
temporal, e.g., \citealp{Athanasopoulos2017-zh, Nystrup2020-ey}: \pkg{thief}; see the notes to Table \ref{tab:pkg-overview} for the package references).  \pkg{bayesRecon}, \pkg{BayesReconPy} support  cross-sectional and temporal reconciliation, but only \pkg{FoReco}, \pkg{FoRecoML} and  \pkg{FoRecoPy} (the \proglang{Python} implementation of \pkg{FoReco}) offer single-source software for cross-sectional, including general linearly constrained series \citep{Girolimetto2024-ft}, temporal and cross-temporal frameworks  \citep{Di_Fonzo2023-dg, Girolimetto2024-jm}.
       
\begin{table}[!htb]
	\centering
    \footnotesize
	\resizebox{\linewidth}{!}{\begin{tabular}{m{0\linewidth}m{0.275\linewidth}|>{\centering\arraybackslash}p{2em}|>{\centering\arraybackslash}p{2em}|>{\centering\arraybackslash}p{2em}|>{\centering\arraybackslash}p{2em}|>{\centering\arraybackslash}p{2em}|>{\centering\arraybackslash}p{2em}|>{\centering\arraybackslash}p{2em}|>{\centering\arraybackslash}p{2em}|>{\centering\arraybackslash}p{2em}|>{\centering\arraybackslash}p{2em}|>{\centering\arraybackslash}p{2em}}
    \toprule
        \multicolumn{2}{l|}{\textbf{}} & 
		\rotatebox[origin=c]{90}{(1)}  & 
		\rotatebox[origin=c]{90}{(2)}  & 
		\rotatebox[origin=c]{90}{(3)}  & 
		\rotatebox[origin=c]{90}{(4)}  & 
		\rotatebox[origin=c]{90}{(5)}  & 
		\rotatebox[origin=c]{90}{(6)}  & 
		\rotatebox[origin=c]{90}{(7)}  & 
		\rotatebox[origin=c]{90}{(8)}  & 
		\rotatebox[origin=c]{90}{(9)}  & 
		\rotatebox[origin=c]{90}{(10)}  & 
		\rotatebox[origin=c]{90}{(11)} \\
        
        \multicolumn{2}{l|}{\textbf{}} & 
		\rotatebox[origin=l]{90}{\tiny -- CRAN}  & 
		\rotatebox[origin=l]{90}{\tiny -- PyPI}  & 
		\rotatebox[origin=l]{90}{\tiny -- CRAN}  & 
		\rotatebox[origin=l]{90}{\tiny -- CRAN}  & 
		\rotatebox[origin=l]{90}{\tiny -- CRAN}  &
        \rotatebox[origin=l]{90}{\begin{minipage}{30pt}
				{\tiny -- CRAN}\\
				{\tiny -- PyPI}
		\end{minipage}} &
		\rotatebox[origin=l]{90}{\tiny -- PyPI}  & 
		\rotatebox[origin=l]{90}{\tiny -- PyPI}  & 
		\rotatebox[origin=l]{90}{\tiny -- PyPI}  & 
		\rotatebox[origin=l]{90}{\tiny }  & 
		\rotatebox[origin=l]{90}{\tiny }  \\
        
        \multicolumn{2}{l|}{\textbf{}} & 
		\rotatebox[origin=l]{90}{\proglang{R}} & 
		\rotatebox[origin=l]{90}{\proglang{Py}} & 
		\rotatebox[origin=l]{90}{\proglang{R}} & 
		\rotatebox[origin=l]{90}{\proglang{R}} & 
		\rotatebox[origin=l]{90}{\proglang{R}} & 
		\rotatebox[origin=l]{90}{\begin{minipage}{10pt}
				\proglang{R}\\
				\proglang{Py}
		\end{minipage}} & 
		\rotatebox[origin=l]{90}{\proglang{Py}} & 
		\rotatebox[origin=l]{90}{\proglang{Py}} & 
		\rotatebox[origin=l]{90}{\proglang{Py}} & 
		\rotatebox[origin=l]{90}{\proglang{R}} &
		\rotatebox[origin=l]{90}{\proglang{R}} \\
		\multicolumn{2}{l|}{\textbf{Functionality}} & 
		\rotatebox[origin=l]{90}{\begin{minipage}{2.5cm}
				\pkg{FoReco}\\
				\pkg{FoRecoML}
		\end{minipage}} & 
		\rotatebox[origin=l]{90}{\begin{minipage}{2.5cm}\pkg{FoRecoPy}\end{minipage}} &
		\rotatebox[origin=l]{90}{\begin{minipage}{2.5cm}\pkg{hts}\end{minipage}} & 
		\rotatebox[origin=l]{90}{\begin{minipage}{2.5cm}\pkg{thief}\end{minipage}} & 
        \rotatebox[origin=l]{90}{\begin{minipage}{2.5cm}
			\pkg{Fable}/\pkg{fabletools}
		\end{minipage}} &
		\rotatebox[origin=l]{90}{\begin{minipage}{2.5cm}
				\pkg{bayesRecon}\\
				\pkg{BayesReconPy}
		\end{minipage}} &
		\rotatebox[origin=l]{90}{\begin{minipage}{2.5cm}\pkg{scikit-hts}\end{minipage}} &
		\rotatebox[origin=l]{90}{\begin{minipage}{2.5cm}\pkg{reconcile}\end{minipage}} &
		\rotatebox[origin=l]{90}{\begin{minipage}{2.5cm}
				\pkg{pyhts}
		\end{minipage}} & 
		\rotatebox[origin=l]{90}{\begin{minipage}{2.5cm}\pkg{ProbReco}\end{minipage}} & 
		\rotatebox[origin=l]{90}{\begin{minipage}{2.5cm}\pkg{gtop}\end{minipage}} \\
		\midrule
		\multicolumn{10}{l}{\textit{Reconciliation Frameworks}}\\
		& Cross-sectional (Hierarchical/grouped)$^a$ & \cmark & \cmark & \cmark & \xmark & \cmark& \cmark & \cmark & \cmark & \cmark & \cmark & \cmark\\
		& Cross-sectional \linebreak (Linearly constrained)$^b$ & \cmark & \cmark & \xmark & \xmark & \xmark& \xmark & \xmark & \xmark & \xmark & \xmark & \xmark\\
		& Temporal$^c$ & \cmark & \cmark & \xmark & \cmark & \xmark& \cmark & \xmark & \xmark & \xmark & \xmark & \xmark\\
		& Cross-temporal$^d$ & \cmark & \xmark & \xmark & \xmark & \xmark& \xmark & \xmark & \xmark & \xmark & \xmark & \xmark\\
		
		\multicolumn{10}{l}{\textit{Reconciliation Approaches}}\\
		& Classical (bottom-up, top-down and midlle-out)$^e$ & \cmark & \xmark & \cmark & \xmark & \cmark & \xmark & \cmark & \xmark & \xmark & \xmark & \xmark\\
		& Regression‑based/Min trace$^f$ & \cmark & \cmark & \cmark & \cmark & \cmark & \xmark & \cmark & \xmark & \cmark & \xmark & \cmark\\
		& Score optimisation$^g$ & \xmark & \xmark & \xmark & \xmark & \xmark & \xmark & \xmark & \cmark & \xmark & \cmark & \xmark\\
		& Bayesian$^h$ & \xmark & \xmark & \xmark & \xmark & \xmark& \cmark & \xmark & \cmark & \xmark & \xmark & \xmark\\
		& Level Conditional Coherent$^i$ & \cmark & \xmark & \xmark & \xmark & \xmark& \xmark & \xmark & \xmark & \xmark & \xmark & \xmark\\
		& Machine-learning$^j$ & \cmark & \xmark & \xmark & \xmark & \xmark& \xmark & \xmark & \xmark & \xmark & \xmark & \xmark\\
		& Probabilistic$^k$ & \cmark & \xmark & \xmark & \xmark & \cmark & \cmark & \xmark & \cmark & \xmark & \cmark & \xmark\\     
		
		\multicolumn{10}{l}{\textit{Additional Functionality}}\\
		& Non-negative constraints$^l$ & \cmark & \cmark & \xmark & \xmark & \xmark& \xmark & \xmark & \xmark & \xmark & \xmark & \xmark\\
		& Immutable forecasts$^m$ & \cmark & \cmark & \xmark & \xmark & \xmark& \xmark & \xmark & \xmark & \xmark & \xmark & \xmark\\
		& Temporal aggregation (sum, average, first/last)$^n$ & \cmark & \cmark & \xmark & \xmark & \xmark& \xmark & \xmark & \xmark & \xmark & \xmark & \xmark\\
        
		\bottomrule
	\end{tabular}}
		\vspace*{-0.75em}
    \begin{flushleft}\footnotesize
    \textbf{Functionality references:}
    \vspace*{-1.25em}
    \begin{multicols}{3}\scriptsize
    \begin{enumerate}[label={$^{\alph*}$}, nosep, leftmargin = *, labelsep=1pt]
        \item \cite{Hyndman2011-fh}.
        \item \cite{Girolimetto2024-ft}.
        \item \cite{Athanasopoulos2017-zh, Nystrup2020-ey}.
        \item \cite{Di_Fonzo2023-dg, Girolimetto2024-jm}.
        \item \cite{Dangerfield1992-ks, Fliedner2001-vx, Athanasopoulos2009-gs}.
        \item \cite{Stone1942-fa, Byron1978-ws, Byron1979-hv, Panagiotelis2021-rh, Wickramasuriya2019}.
        \item \cite{Panagiotelis2023-se}.
        \item \cite{Corani2021-de, Corani2024-mv, Zambon2023-jn, Zambon2024-pw}.
        \item \cite{Hollyman2021-zq, Di_Fonzo2024-ym}.
        \item \cite{Spiliotis2021-uo, Rombouts2025-ym}.
        \item \cite{Panagiotelis2023-se, Wickramasuriya2024-yv}
         \item \cite{Wickramasuriya2020-zk, Di_Fonzo2023-ae, nn2025}.
        \item \cite{Zhang2023-yj}.
        \item \cite{Chow1971-hs, zellner1971study}.
    \end{enumerate}
    \end{multicols}
    \vspace*{-1em}
    \textbf{Packages references:}
    \vspace*{-1.25em}
    \begin{multicols}{3}\scriptsize
    \begin{enumerate}[label={$^{(\arabic*)}$}, nosep, leftmargin = *, labelsep=1pt]
        \item \cite{FoReco, FoRecoML}.
        \item \cite{FoRecoPy}.
        \item \cite{hts}.
        \item \cite{thief}.
        \item \cite{fable, fabletools}.
        \item \cite{bayesRecon, BayesReconPy}.
        \item \cite{scikit-hts}.
        \item \cite{reconcile}.
        \item \cite{pyhts}.
        \item \cite{ProbReco}.
        \item \cite{gtop}.
    \end{enumerate}
    \end{multicols}
    \end{flushleft}
    \vspace*{-1em}
    \caption{\label{tab:pkg-overview}Overview of existing \proglang{R} and \proglang{Python} packages (columns) and the main forecast reconciliation functionalities they support (rows).} 
\end{table}

Regarding forecast reconciliation approaches, Table \ref{tab:pkg-overview} reveals that
\pkg{hts}, \pkg{scikit-hts}, \pkg{Fable}, \pkg{fabletools} offer popular classical  \citep{Dangerfield1992-ks, Fliedner2001-vx} and regression-based \citep{Stone1942-fa, Byron1978-ws, Byron1979-hv, Wickramasuriya2019, Panagiotelis2021-rh, Panagiotelis2023-se, Wickramasuriya2024-yv} linear reconciliation approaches, with the former two packages focusing on point forecasting and the latter also on probabilistic forecasting.
Specialized software exists for
regression-based approaches (\pkg{thief}, \pkg{gtop} and \pkg{pyhts}) or for
probabilistic forecasting  including score optimization (\citealp{Panagiotelis2023-se}; \pkg{ProbReco}), Bayesian methods (\citealp{Corani2021-de, Corani2024-mv, Zambon2023-jn, Zambon2024-pw}; \pkg{bayesRecon}, \pkg{BayesReconPy}) and approaches that address both  (\pkg{reconcile}).
\pkg{FoReco} and \pkg{FoRecoML} offer a more comprehensive alternative, providing functionality for a wide range of reconciliation approaches within a standardized and unified framework.
Moreover, they are the only packages that offer specialized functionality for 
level conditional reconciled forecasts \citep{Hollyman2021-zq, Di_Fonzo2024-ym}, and accommodate
additional features such as non-negative forecast reconciliation \citep{Wickramasuriya2020-zk, Di_Fonzo2023-ae, nn2025}, immutable reconciliation that preserves forecasts for a pre-specified subset of variables unchanged \citep{Zhang2023-yj}, or various temporal aggregation schemes -- for example, summation for flow variables or averaging/sampling at specific points for stock variables in economics \citep{zellner1971study, Chow1971-hs}.
Finally, \pkg{ForecoML}
is the only package providing non-linear reconciliation  via machine learning (ML)
\citep{Spiliotis2021-uo, Rombouts2025-ym}.

\pkg{Foreco} and \pkg{ForecoML} are thus unique in providing a single-source software solution -- a comprehensive toolbox encompassing a wide range of reconciliation methods across all reconciliation frameworks. 
To accomplish this philosophy, we provide a unified software design, consistent not only across all supported reconciliation frameworks and methods but also across \pkg{Foreco} and \pkg{ForecoML}, thereby facilitating ease of use for the end user.
All reconciliation functions take base forecast as main input, via the universal argument \code{base}, and return reconciled forecasts of the same dimension.
Both packages share a consistent syntax across all reconciliation functions: each function's pre-fix indicates the reconciliation framework (\code{cs*} for cross-sectional, \code{te*} for temporal and \code{ct*} for cross-temporal) and the suffix indicates the chosen reconciliation method.
The syntax of framework-specific arguments is  shared across all reconciliation functions, promoting ease of use and control. In addition, both packages return an object of the same class \code{foreco} across all reconciliation functions, which provides common \code{print()}, \code{summary()}, \code{plot()} and \code{components()} methods for inspecting the reconciled forecasts regardless of the framework or method used.
All reconciliation functions are fully user-configurable with respect to the reconciliation approach, with the option to rely on sensible defaults for near-automatic implementation.

The packages \pkg{FoReco} and \pkg{FoRecoML} are deliberately kept separate. This separation is desirable because \pkg{FoRecoML} supports ML-based reconciliation, which in turn requires additional dependencies on other \proglang{R} packages, namely 
\pkg{randomForest} \citep{randomForest}, \pkg{lightgbm} \citep{lightgbm}, \pkg{xgboost} \citep{xgboost}, and \pkg{mlr3} \citep{mlr3}, implementing these methods. In addition, we include more learners and tuning options for model parameters through \pkg{mlr3tuning} \citep{mlr3tuning}, \pkg{mlr3learners} \citep{mlr3learners}, and \pkg{paradox} \citep{paradox}. 
In contrast, \pkg{FoReco} does not require these dependencies, and to preserve a lightweight minimally dependent design, it is kept separate of  \pkg{FoRecoML}. Nonetheless, the latter adheres to the same design principles as \pkg{FoReco}, which explains its dependency on  \pkg{FoReco} as it inherits time series processing functionalities from the latter.

Note that  \pkg{FoReco} does not include functionality for score optimisation (see Table \ref{tab:pkg-overview}); this reconciliation method is typically outperformed by regression-based reconciliation approaches, and is typically more computationally demanding than the latter \citep{Panagiotelis2023-se}.
Furthermore, neither \pkg{FoReco} nor \pkg{FoRecoML} provide functionality for Bayesian reconciliation, which typically requires fundamentally different formulations and implementations. Several specialized software packages  already exist for Bayesian reconciliation (e.g., \pkg{bayesRecon} and \pkg{BayesReconPy}); we refer interested users to those.

\pkg{FoReco} and \pkg{FoRecoML} are available from the Comprehensive \proglang{R} Archive Network (CRAN), with \pkg{FoReco} at \texttt{v1.3.0} and \pkg{ForecoML} at \texttt{v1.1.0}. The latest (development) versions are  
available on GitHub at \url{https://github.com/danigiro/FoReco} and \url{https://github.com/danigiro/FoRecoML} respectively.
Version \texttt{1.3.0} of \pkg{FoReco}, version \texttt{1.1.0} of \pkg{FoRecoML}, and version \texttt{4.6.0} of \proglang{R} were used in this paper.

The remainder of the paper is structured as follows.
Section \ref{sec:ctreco} reviews classical, regression-based and machine-learning based reconciliation methods for cross-sectional, temporal and cross-temporal frameworks. 
Section \ref{sec:packages} presents the packages' implementation.
Section \ref{sec:illustration-with-data} uses  applications on cross-sectional, temporal and cross-temporal reconciliation to illustrate the usage and usefulness of \pkg{Foreco} and \pkg{ForecoML}.
Section \ref{sec:conclusion} concludes.

\section[Forecast reconciliation]{Forecast reconciliation} \label{sec:ctreco}
We review linearly constrained multiple time series across cross-sectional, temporal and cross-temporal frameworks in Section \ref{subsec:frameworks}.
Section \ref{subsec:methods} reviews widely used
reconciliation approaches, namely linear reconciliation approaches supported in \pkg{FoReco},
and non-linear reconciliation approaches supported in \pkg{FoRecoML}.
We hereby  provide a concise, practical introduction focused on essential concepts that practitioners need to implement forecast reconciliation methods. 
For a more extensive recent overview of forecast reconciliation, we refer the  reader to \cite{athanasopoulos2023review}.

\subsection{Reconciliation frameworks} \label{subsec:frameworks}
We review the key concepts for cross-sectional, temporal and cross-temporal frameworks.

\begin{figure}[!tbh]
	\captionsetup[subfigure]{justification=centering}
\begin{subfigure}[t]{.5\linewidth}
\centering
\includegraphics[width=0.75\linewidth, keepaspectratio]{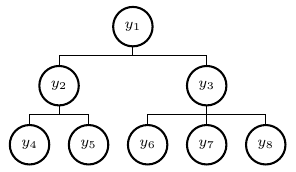} \\[0.5em]
\textbf{\small Matrix form:}
{\scriptsize
$$\arraycolsep=1pt
\bm{A}_{cs} = \begin{bmatrix}
1 & 1 & 1 & 1 & 1 \\
1 & 1 & 0 & 0 & 0 \\
0 & 0 & 1 & 1 & 1 \\
\end{bmatrix}, \; \bm{C}_{cs} = \begin{bmatrix}
1 & 0 & 0 & -1 & -1 & -1 & -1 & -1 \\
0 & 1 & 0 & -1 & -1 & 0 & 0 & 0 \\
0 & 0 & 1 & 0 & 0 & -1 & -1 & -1 \\
\end{bmatrix}$$}
\vspace*{-1em}
\caption{Cross-Sectional Framework: $y_1 = y_2 + y_3$, $y_2 = y_4 + y_5$ and $y_3 = y_6 + y_7 + y_8$} \label{fig:toy:cs}
\end{subfigure}
\begin{subfigure}[t]{.5\linewidth}
\centering
\includegraphics[width=0.75\linewidth, keepaspectratio]{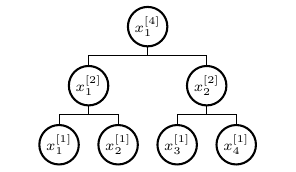} \\[0.5em]
\textbf{\small Matrix form ($m = 4$, $\mathcal{K} =\{4, 2, 1\}$): }
{\scriptsize
$$\arraycolsep=1pt
\bm{A}_{te} = \begin{bmatrix}
1 & 1 & 1 & 1 \\
1 & 1 & 0 & 0 \\
0 & 0 & 1 & 1 \\
\end{bmatrix}, \; \bm{C}_{te} = \begin{bmatrix}
1 & 0 & 0 & -1 & -1 & -1 & -1 \\
0 & 1 & 0 & -1 & -1 & 0 & 0 \\
0 & 0 & 1 & 0 & 0 & -1 & -1 \\
\end{bmatrix}$$}
\vspace*{-1em}
\caption{Temporal Framework: quarterly over semi-annual to annual.} \label{fig:toy:te}
\end{subfigure}
\vskip1em
\begin{subfigure}{\linewidth}
\centering
\includegraphics[width=\linewidth, keepaspectratio]{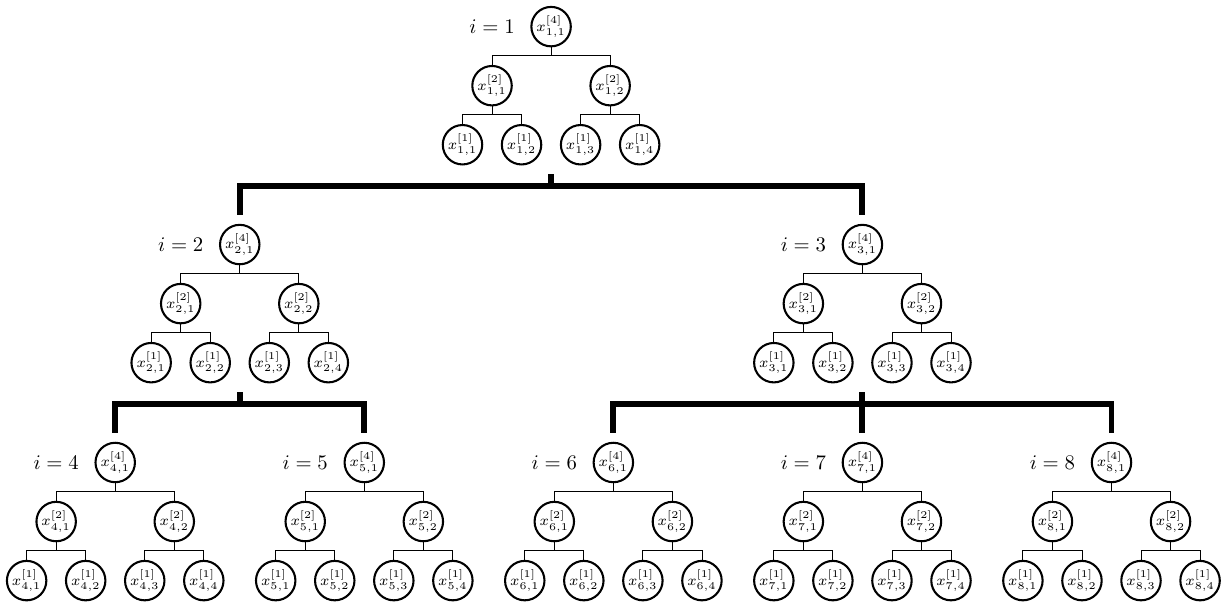} 
\caption{Cross-Temporal Framework: quarterly over semi-annual to annual data ($m = 4$) with cross-sectional constraints such that $x_{1, j}^{[k]} = x_{2, j}^{[k]} + x_{3, j}^{[k]}$, $x_{2, j}^{[k]} = x_{4, j}^{[k]} + x_{5, j}^{[k]}$, and $x_{3, j}^{[k]} = x_{6, j}^{[k]} + x_{7, j}^{[k]} + x_{8, j}^{[k]}$ with $j = 1,...,\frac{4}{k}$ and $k \in \{4,2,1\}$} \label{fig:toy:ct}
\end{subfigure}
\caption{Toy examples for cross-sectional, temporal and cross-temporal frameworks.} \label{fig:toy}
\end{figure}

\paragraph{Cross-sectional framework}
Let $\bm{y}_t = [ y_{1,t}\; \ldots\; y_{n,t}]^\top$ denote an $n$-variate linearly constrained time series observed at time $t$, whose components are subject to $n_u$ linear constraints. A common special case are hierarchical or grouped time series, where $\bm{y}_t$ consists of $n_b$ \textit{bottom-level} series and $n_u$ \textit{upper-level} series, with $n= n_u + n_b$.
Figure \ref{fig:toy:cs} shows a cross-sectional hierarchical structure as a toy example with $n=8$  series of which $n_b=5$ bottom-level series.
More generally, the constraints among the series can be expressed as a linear systems with $n_u$ equations and $n$ variables given by
\begin{equation}
\bm{C}_{cs} \bm{y}_t = \bm{0} \label{eq:cs-C}
\end{equation}
where $\bm{C}_{cs}$ is the $n_u \times n$-dimensional cross-sectional \textit{constraint matrix}.
Alternatively, the $n_u$-dimensional vector of upper-level series $\bm{u}_t$ can be expressed in terms of the $n_u\times n_b$-dimensional \textit{aggregation matrix} $\bm{A}_{cs}$ and the $n_b$-dimensional vector of bottom-level series $\bm{b}_t$, namely
\begin{equation}
\bm{u}_t = \bm{A}_{cs} \bm{b}_t. \label{eq:cs-A}
\end{equation}
Hence, either the constraint matrix $\bm{C}_{cs}$ in Equation~\ref{eq:cs-C} or the aggregation matrix $\bm{A}_{cs} $ in Equation~\ref{eq:cs-A} fully characterize the linearly constrained system of series in the cross-sectional hierarchy.
The two representations are related by $\bm{C}_{cs} = [\bm{I}_{n_u}\; -\bm{A}_{cs}]$, and $\bm{A}_{cs}$ can be obtained from a suitable transformation of $\bm{C}_{cs}$, as discussed in \cite{Girolimetto2024-ft}. Users may therefore specify either $\bm{A}_{cs}$ or $\bm{C}_{cs}$ when performing cross-sectional reconciliation with \pkg{FoReco} or \pkg{FoRecoML}.

\paragraph{Temporal framework}
Temporal reconciliation, proposed by \cite{Athanasopoulos2017-zh}, relates an original high-frequency series, typically denoted by $y_t$, to various temporally aggregated series, typically denoted by $x_j$.   
Let $m$ denote the \textit{maximum aggregation order}, namely the number of observations to sum over when going from the highest to the lowest sampling frequency in the hierarchy.
For $p$ levels in the temporal hierarchy, let $\mathcal{K} = \{k_p, \ldots, k_1\}$ denote the aggregation levels (in descending order) for all $p$ levels, with $k_1=1$ by convention for the bottom-level in the temporal hierarchy (no aggregation) and $k_p = m$ for the top-level in the hierarchy. 
Given the set of \textit{aggregation orders}, to be specified as input in \pkg{FoReco} and \pkg{FoRecoML},
the non-overlapping temporally
aggregated series can then be constructed for each level $k= k_1, \ldots, k_p$, as given by
\begin{equation}
x_{j}^{[k]}=\sum_{t=(j-1)k + 1}^{jk} y_{t}, \qquad j=1,\ldots,m/k. \label{eq:ct}
\end{equation}
Figure \ref{fig:toy:te} shows a temporal framework where a quarterly series ($k_1=1$) is first aggregated semi-annually ($k_2=2)$ then annually ($k_3=m=4)$. 

Finally, while most studies on temporal reconciliation focus on summing high-frequency to obtain low-frequency series, as in Equation~\ref{eq:ct}, the type of temporal aggregation can easily be adapted to use averages or specific values (e.g., the first or last observation) depending on the nature of the variables, such as flow versus stock variables in economics \citep{zellner1971study, Chow1971-hs}.

\paragraph{Cross-temporal framework}
Cross-sectional and temporal constraints can occur jointly, resulting in a cross-temporal framework as discussed in \cite{Di_Fonzo2023-dg}.
Then $\bm{y}_t$ denotes the $n$-dimensional vector of all cross-sectional series at the highest sampling frequency (i.e., the bottom level in the temporal hierarchy). 
Equations~\ref{eq:cs-C} and~\ref{eq:cs-A} still contain the information on how to cross-sectionally aggregate the series regardless of the temporal aggregation level, though the temporal observation index now varies with the temporal aggregation level. 
Equation~\ref{eq:ct}  simply needs to be augmented with an additional subindex $i= 1, \ldots, n$ such that, for each $i$th cross-sectional series, all temporally aggregated series $x_{i,j}^{[k]}$ (for $k=k_1, \ldots, k_p$) can be constructed from the bottom-level series $y_{i,t}$ in the temporal hierarchy.
Figure \ref{fig:toy:ct} shows a toy example with a cross-temporal hierarchy.

Any reconciliation approach starts from a set of \textit{base forecasts} for all series subject to linear constraints. Without loss of generality, we present the essential notation for the cross-temporal case, which includes the cross-sectional and temporal ones.

Let   $\widehat{\bm{X}}$ be the $n\times q$ matrix that collects all base forecasts for forecast horizon $H$ (for the most temporally aggregated series). The forecasts for the $n$ cross-sectional units are hereby contained in the rows, the forecasts across the different temporal aggregation levels in the columns where
$q=\sum_{k \in \mathcal{K}} mH/k$ and $\hat{\bm{x}}_i^{[k]}$ denotes the $mH/k$-dimensional row vector collecting all forecasts at temporal aggregation level $k$ for cross-sectional series $i$.
The base forecasts matrix is then given by
\begin{equation}
\label{eq:bf_mat}
\widehat{\bm{X}} = 
\begin{bmatrix}
\hat{\bm{x}}_{1}^{[k_p]} & \hat{\bm{x}}_{1}^{[k_p -1]} & \hdots & \hat{\bm{x}}_{1}^{[k_1]} \\
\hat{\bm{x}}_{2}^{[k_p]} & \hat{\bm{x}}_{2}^{[k_p -1]} & \hdots & \hat{\bm{x}}_{2}^{[k_1]} \\
\vdots & \vdots & \ddots & \vdots \\
\hat{\bm{x}}_{n}^{[k_p]} & \hat{\bm{x}}_{n}^{[k_p -1]} & \hdots & \hat{\bm{x}}_{n}^{[k_1]} \nonumber
\end{bmatrix}.
\end{equation}
The base forecasts, generally, do not satisfy the cross-sectional and temporal constraints. 
The objective of any reconciliation approach is therefore to revise the base forecasts to obtain \textit{reconciled forecasts} collected in a new matrix $\widetilde{\bf{X}}$ that are coherent; they satisfy the cross-sectional and temporal constraints.
Any reconciliation function in \pkg{FoReco} and \pkg{FoRecoML} therefore takes the base forecasts  $\widehat{\bf{X}}$ as key input and returns the reconciled forecasts  $\widetilde{\bf{X}}$ as output.
Next, we review popular reconciliation approaches supported in \pkg{FoReco} and \pkg{FoRecoML}. 

\subsection{Reconciliation approaches} \label{subsec:methods}
\pkg{FoReco} supports functionality for classical, regression-based and probabilistic forecast reconciliation.
\pkg{FoRecoML} supports functionality for ML-based reconciliation.

\paragraph{Classical reconciliation}
Classical "single-level" reconciliation approaches select one level of aggregation for which forecasts are generated and then linearly combined to obtain a  coherent forecasts for all levels in the cross-temporal hierarchy.
This includes top-down, bottom-up and middle-out approaches.

Top-down approaches only require base forecasts for the most aggregated (top) level in the cross-sectional/temporal/cross-temporal hierarchy and then disaggregate these forecasts down  according to a proportional weighting scheme to be supplied by the user.
The top-level forecast thus remains unchanged.
Bottom-up approaches only require base forecasts at the bottom of the hierarchy and then appropriately aggregate these up.
Middle-out approaches require forecasts at some chosen intermediate level $\ell$ in the hierarchy and then combine
a bottom-up approach for all levels above level $\ell$ with 
a top-down approach (with user-specified weights)  for all levels below level $\ell$ in the hierarchy. 

Classical single-level approaches are, however, restricted to using information from a single level of the hierarchy, potentially overlooking valuable information from other levels. In contrast, regression-based reconciliation methods incorporate information from all hierarchical levels when revising the base forecasts.

\paragraph{Regression-based reconciliation}
Regression-based reconciliation approaches include least-squares based reconciliation and  level conditional coherent reconciliation.

Least-squares based reconciliation uses either a projection approach \citep{Byron1978-ws, Byron1979-hv} or an equivalent structural approach \citep{Hyndman2011-fh} to obtain the reconciled forecasts. 
The
 reconciled forecasts, using notation from the projection approach, are 
\begin{equation}
\widetilde{\bm{x}} = \bm{M}_\text{ct}\widehat{\bm{x}}, \nonumber
\end{equation}
with $\widetilde{\bm{x}} = \text{vec}(\widetilde{\bm{X}}^\top)$, $\widehat{\bm{x}} = \text{vec}(\widehat{\bm{X}}^\top)$, and $\bm{M}_\text{ct}$ is the cross-temporal projection matrix 
given by
\begin{equation}
 \bm{M}_\text{ct}  = \bm{I} - \bm{\Omega}_\text{ct}\bm{C}_\text{ct}^\top\left(\bm{C}_\text{ct} \bm{\Omega}_\text{ct} \bm{C}_\text{ct}^\top \right)^{-1}\bm{C}_\text{ct}, 
 \label{eq:reg:M}
\end{equation}
with $\bm{\Omega}_\text{ct}$ a positive definite covariance matrix of base forecast errors, and  $\bm{C}_\text{ct}$ the constraint matrix encoding all constraints across the whole cross-temporal framework, see e.g., \cite{Di_Fonzo2023-dg} for details. Similar expressions for the projection matrix hold for a pure cross-sectional system (with $\bm{\Omega}_\text{cs}$ and $\bm{C}_\text{cs}$) or temporal hierarchy (with $\bm{\Omega}_\text{te}$ and $\bm{C}_\text{te}$) .

The covariance matrix $\bm{\Omega}_\text{ct}$ forms an important component in Equation~\ref{eq:reg:M} as it allows to take the quality of the initial base forecasts into account; forecasts with smaller error variance get more weight. 
Different choices to approximate $\bm{\Omega}_\text{ct}$ are available in the literature and lead to different reconciliation solutions, see Table~\ref{tab:Omega} for an overview; all are supported in \pkg{FoReco}.

\begin{table}[t]
	\centering
	\resizebox{\linewidth}{!}{
		\begin{tabular}{m{0.27\linewidth}|p{7em}|p{8.5em}|p{10em}}
			\toprule
			&  \multicolumn{3}{c}{\textbf{Framework}}\\
			\textbf{Description} & \multicolumn{1}{c}{Cross-sectional} &  \multicolumn{1}{c}{Temporal} &  \multicolumn{1}{c}{Cross-temporal}\\
			\midrule
			\textit{Ordinary LS 
            } & 
			\begin{minipage}[t]{\linewidth}%
				\begin{itemize}[nosep, leftmargin=!, labelwidth=0.75em]%
					\item[$^{1}$:] \code{ols}
				\end{itemize}%
			\end{minipage} & 
			\begin{minipage}[t]{\linewidth}%
				\begin{itemize}[nosep, leftmargin=!, labelwidth=0.75em]%
					\item[$^{2}$:] \code{ols}
				\end{itemize}%
			\end{minipage}  &
			\begin{minipage}[t]{\linewidth}%
				\begin{itemize}[nosep, leftmargin=!, labelwidth=0.75em]%
					\item[$^{3}$:] \code{ols}
				\end{itemize}%
			\end{minipage}  \\
			\addlinespace
            \textit{Weighted LS 
            } &
			\begin{minipage}[t]{\linewidth}%
				\begin{itemize}[nosep, leftmargin=!, labelwidth=0.75em]%
					\item[$^{2}$:] \code{str}
				\end{itemize}%
			\end{minipage} & 
			\begin{minipage}[t]{\linewidth}%
				\begin{itemize}[nosep, leftmargin=!, labelwidth=0.75em]%
					\item[$^{2}$:] \code{str}
				\end{itemize}%
			\end{minipage}  &
			\begin{minipage}[t]{\linewidth}%
				\begin{itemize}[nosep, leftmargin=!, labelwidth=0.75em]%
					\item[$^{3}$:] \code{str}
					\item[$^{4}$:] \code{csstr}, \code{testr}
				\end{itemize}%
			\end{minipage}  \\
			\addlinespace
			\textit{Weighted LS with \code{res}} & 
			\begin{minipage}[t]{\linewidth}%
				\begin{itemize}[nosep, leftmargin=!, labelwidth=0.75em]%
					\item[$^{5}$:] \code{wls}
				\end{itemize}%
			\end{minipage} & 
			\begin{minipage}[t]{\linewidth}%
				\begin{itemize}[nosep, leftmargin=!, labelwidth=0.75em]%
					\item[$^{2}$:] \code{wlsh}, \code{wlsv} 
				\end{itemize}%
			\end{minipage}  &
			\begin{minipage}[t]{\linewidth}%
				\begin{itemize}[nosep, leftmargin=!, labelwidth=0.75em]%
					\item[$^{3}$:] \code{wlsh}, \code{wlsv}
				\end{itemize}%
			\end{minipage}  \\
			\addlinespace
            \textit{Generalized LS with \code{res}} & 
			\begin{minipage}[t]{\linewidth}%
				\begin{itemize}[nosep, leftmargin=!, labelwidth=0.75em]%
					\item[$^{6}$:] \code{sam}, \code{shr}
					\item[$^{7}$:] \code{oasd}
				\end{itemize}%
			\end{minipage} & 
			\begin{minipage}[t]{\linewidth} \raggedleft
				\begin{itemize}[nosep, leftmargin=!, labelwidth=0.75em]%
					\item[$^{2}$:] \code{sam}, \code{shr}
					\item[$^{8}$:] \code{acov}, \code{strar1}, \\ \code{sar1}, \code{har1}
				\end{itemize}%
			\end{minipage}  &
			\begin{minipage}[t]{\linewidth}%
				\begin{itemize}[nosep, leftmargin=!, labelwidth=0.75em]%
					\item[$^{4}$:] \code{sam}, \code{shr}, \\ \code{acov}, \code{bdsam}, \code{bdshr}
					\item[$^{9}$:] \code{Ssam}, \code{Sshr}
					\item[$^{10}$:] \code{bsam}, \code{bshr}, \code{hsam}, \\ \code{hshr}, \code{hbsam}, \code{hbshr},  
				\end{itemize}%
			\end{minipage}  \\
			\bottomrule
		\end{tabular}
	}
		\vspace*{-0.75em}
	\begin{flushleft}\footnotesize
		\textbf{Covariances references:}
		\vspace*{-1.25em}
		\begin{multicols}{3}\scriptsize
			\begin{enumerate}[label={$^{(\arabic*)}$}, nosep, leftmargin = *]
				\item \cite{Hyndman2011-fh}.
				\item \cite{Athanasopoulos2017-zh}.
				\item \cite{Di_Fonzo2023-dg}.
				\item \cite{Girolimetto2025-smap}.
				\item \cite{Hyndman2016-wj}.
				\item \cite{Wickramasuriya2019}.
				\item \cite{Ando2023-am}.
				\item \cite{Nystrup2020-ey}
				\item \cite{Bisaglia2025-gg}.
				\item \cite{Girolimetto2024-jm}.
			\end{enumerate}
		\end{multicols}
	\end{flushleft}
\vspace*{-1em}
\caption{\label{tab:Omega}Choices of the base forecasts covariance matrix for least-squares (LS) reconciliation in \pkg{FoReco}. Abbreviations correspond to the function arguments introduced in Section \ref{sec:packages}. \\ Note: “\textit{with \code{res}}” means that the corresponding methods additionally require 
    in-sample residuals or validation errors \code{res} to be provided by the user. 
    }
\vspace*{-1em}
\end{table}

The first, most simple option is to set $\bm{\Omega}_\text{ct} = \bm{I}$, which boils down to using ordinary least squares (LS). All series are  treated as equally reliable and covariances are ignored (\code{ols} in Table~\ref{tab:Omega}; see \citealp{Hyndman2011-fh}, \citealp{Athanasopoulos2017-zh}, \citealp{Di_Fonzo2023-dg} respectively for cross-sectional, temporal and cross-temporal frameworks). 
To account for the quality of the base forecasts,  weighted or generalized LS methods can be used.

Weighted LS methods use a diagonal $\bm{\Omega}_\text{ct}$; they only require variance estimates (no covariances) which makes the corresponding reconciliation solutions typically fast and stable.  The diagonal elements can be "structural", as derived  from the aggregation structure of the series (\code{str} in Table~\ref{tab:Omega}; see \citealp{Athanasopoulos2017-zh, Di_Fonzo2023-dg});  for cross-temporal frameworks, one can opt for using only the cross-sectional structure $\bm{\Omega}_\text{ct} = \bm{\Omega}_\text{cs} \otimes \bm{I}_{\sum_{k \in \mathcal{K}} m/k}$ (\code{csstr}) 
or the temporal structure $\bm{\Omega}_\text{ct} = \bm{I}_{n} \otimes \bm{\Omega}_\text{te}$ (\code{testr}), where $\bm{\Omega}_\text{cs}$ and $\bm{\Omega}_\text{te}$ are respectively the cross-sectional and temporal structural covariance matrix, see \cite{Girolimetto2025-smap}.
Alternatively, the variances can be estimated based on in-sample residuals or validation errors, which need to be additionally provided by the user.
For cross-sectional frameworks, \code{wls} (\citealp{Hyndman2016-wj}) uses the forecast error variances as diagonal entries, whereas for temporal or cross-temporal frameworks, \code{wlsh} (hierarchy variance scaling) allows different variance estimates across and within temporal levels for each series, and \code{wlsv} (series variance scaling) assumes a common variance within each temporal level to simplify estimation and increase sample size (\citealp{Athanasopoulos2017-zh, Di_Fonzo2023-dg}).

Generalized LS methods use a full-blown $\bm{\Omega}_\text{ct}$, making them the most flexible but the additional need for covariance estimators may make them ill-conditioned in high-dimensional or small-sample settings. Then, shrinkage are valid alternatives and generally recommended.
For cross-sectional frameworks, one can 
use the sample covariance matrix of all the base forecasts (\code{sam} in Table~\ref{tab:Omega}) or a shrunk alternative that either assumes knowledge of the true "oracle"  covariance structure (\code{oasd}, \citealp{Ando2023-am})  or not (\code{shr, \citealp{Wickramasuriya2019}}).
For temporal and cross-temporal frameworks, \code{sam} and \code{shr} are also available. 
Additionally, for temporal frameworks, 
more options are available that exploit the temporal dynamics of each series \citep{Hyndman2016-wj}, either using 
its autocovariances (\code{acov}), a structural AR(1) (\code{strar1}), a series-level AR(1) (\code{sar1}), or a  hierarchical AR(1) (\code{har1}).
For cross-temporal frameworks, the autocovariance structure can also be exploited (\code{acov}), and several adaptations of the sample covariance matrix and its shrunk variant are available \citep{Di_Fonzo2023-dg, Bisaglia2025-gg}:
block-diagonal versions assuming temporal (\code{bdsam}, \code{bdshr}; \citealp{Di_Fonzo2023-dg}) and cross-sectional (\code{Ssam}, \code{Sshr}; \citealp{Bisaglia2025-gg}) uncorrelation, ridge regularized covariance matrix \citep{Girolimetto2024-jm} estimated using bottom-level series only (\code{bsam}, \code{bshr}), 
all series at high frequency (\code{hsam}, \code{hshr}), or high-frequency bottom-level series (\code{hbsam}, \code{hbshr}).

Finally, Level Conditional Coherent (LCC) reconciliation, originally introduced by \cite{Hollyman2021-zq} and later extended by \cite{Di_Fonzo2024-ym}, adds an additional constraint to least-squares based reconciliation by requiring that the base forecasts at a chosen level of the hierarchy remain unchanged or "immutable".
Different LCC reconciliations are available depending on the number of levels in the hierarchy, and they can be combined through simple averaging -- known as the Combined Conditional Coherent reconciliation -- which also incorporates the bottom-up reconciled forecasts. As an alternative to the immutable (exogenous) constraint, \cite{Di_Fonzo2024-ym} propose an endogenous formulation in which the upper and bottom levels are jointly revised within each sub-hierarchy.

\paragraph{Probabilistic reconciliation}
All reconciliation forecasts discussed above deliver reconciled point forecasts.
Distributional or probabilistic forecasting has first been introduced for cross-sectional hierarchies by \cite{Panagiotelis2023-se} and \cite{Wickramasuriya2024-yv}, and later extended to temporal and cross-temporal hierarchies by \cite{Girolimetto2024-jm}. Two main classes of reconciliation are Gaussian-based and sample-based probabilistic reconciliation; both are supported in \pkg{FoReco}.

Gaussian-based probabilistic reconciliation forms an immediate natural probabilistic extension of the least-squares based reconciliation approaches.
Assuming the base forecasts $ \widehat{\bm{x}}$ follow a normal distribution with mean $\bm{\mu}$ and variance $\bm{\Sigma}$, the revised forecasts $\widetilde{\bm{x}}$  are also normally distributed,  namely
        $$
        \widehat{\bm{x}} \sim N(\bm{\mu}, \bm{\Sigma}) \Rightarrow \widetilde{\bm{x}}\sim N(\bm{M}\bm{\mu}, \bm{M}\bm{\Sigma}\bm{M}^\top)
        $$
where $\bm{M}$ is the projection matrix of the cross-sectional, temporal or cross-temporal framework; e.g., $\bm{M}_{ct}$ in Equation~\ref{eq:reg:M} for cross-temporal frameworks.  
The covariance matrix $\bm{\Sigma}$ can  be  taken the same as 
the covariance matrix of the forecast errors (i.e., $\bm{\Sigma} = \bm{\Omega}$, for instance $\bm{\Omega}_{ct}$ in Equation~\ref{eq:reg:M}), then 
$\bm{M}\bm{\Sigma}\bm{M}^\top = \bm{M}\bm{\Sigma}$.

More general sample-based approaches are also available. Given a sample of base forecasts, reconciling each sample individually yields a corresponding set of reconciled forecast samples.
To reconcile each sample, any of the reconciliation approaches discussed above (classical or regression-based) may be used.

\paragraph{ML-based reconciliation}
Linear reconciliation approaches adjust base forecasts using linear transformations to ensure that aggregation or hierarchical constraints are satisfied, typically relying on ordinary, weighted, or generalized least squares frameworks to minimize reconciliation errors. These methods are computationally efficient and often provide closed-form solutions, making them straightforward to implement.
In contrast, non-linear reconciliation approaches,  first introduced by \cite{Spiliotis2021-uo} for cross-sectional frameworks and later extended by \cite{Rombouts2025-ym} to cross-temporal frameworks, generate reconciled forecasts directly and automatically through ML models. These approaches are designed to minimize the forecast combination errors at the bottom level of the hierarchy and then use bottom-up aggregation to ensure that the resulting forecasts are coherent across the whole cross-sectional, temporal or cross-temporal framework.
They are generally more computationally demanding than linear reconciliation methods, as no closed-form solutions exist and the ML models must be trained.

The idea behind ML-based reconciliation is to model the relationship between each bottom-level series  of the cross-sectional, temporal or cross-temporal framework and the  base forecasts across the entire framework.
To this end, (out-of-sample) base forecasts for all series need to be obtained on an validation sample $\mathcal{V}$. 
The ML-model can then be generally written as 
\begin{align}
	\bm{y}_{b,\mathcal{V}} = f_{b}(\widehat{\bm{X}}_{b,\mathcal{V}}) + \bm{\varepsilon}_{b,\mathcal{V}}, \label{ML-model}
\end{align}
where $\bm{y}_{b,\mathcal{V}}$ contains the actual values of the bottom-level series on the validation sample $\mathcal{V}$, $\widehat{\bm{X}}_{b,\mathcal{V}}$ contains in its columns the base forecasts for all series in the cross-sectional, temporal or cross-temporal framework on the validation sample $\mathcal{V}$, and $f_b(\cdot)$ denotes the forecast function to be trained for bottom-level series $\bm{y}_{b,\mathcal{V}}$. 
The forecast function $f_b(\cdot)$ in Equation~\ref{ML-model} needs to be estimated for each bottom level series separately and using a ML model chosen by the user (e.g., random forest or XGBoost). By using machine learning, nonlinear relationships among the base forecasts of all series are automatically captured when combining them to revise the forecasts of each bottom-level series.
The trained forecast function can then be used to produce coherent forecasts on a chosen test sample: 
out-of-sample base forecasts for all series in the hierarchy on the test sample are used as inputs to the trained forecast function $\hat{f}_b(\cdot)$ to generate revised forecasts for the bottom-level series. Finally, these revised bottom-level forecasts are aggregated using a simple bottom-up approach to produce coherent forecasts across the entire hierarchy.

\section[FoReco and FoRecoML]{\pkg{FoReco} and \pkg{FoRecoML}} \label{sec:packages}
The packages \pkg{FoReco} and \pkg{FoRecoML} offer post-forecasting functionality to improve the accuracy and coherency of forecasts for systems of linearly constrained series. 
Their core premise is that base forecasts are supplied by the user; given these,  \pkg{FoReco} and \pkg{FoRecoML}  reconcile them to obtain coherency across the cross-sectional, temporal or cross-temporal framework.

\begin{figure}[p]
    \centering
    \begin{subfigure}{\linewidth}
    \centering
    \includegraphics[width=0.9\linewidth]{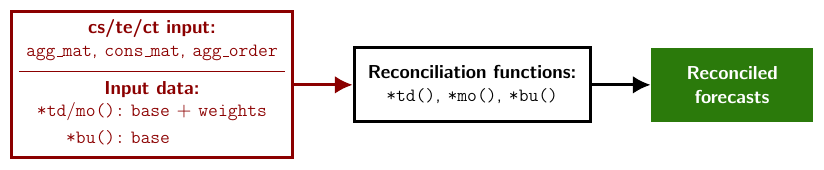}
    \caption{Classical reconciliation workflow} \label{fig:workflow:class}
    \end{subfigure}
    \begin{subfigure}{\linewidth}
    \centering
    \includegraphics[width=0.9\linewidth]{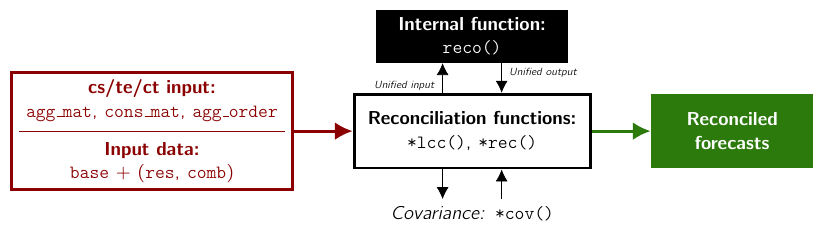}
    \caption{Regression-based reconciliation workflow} \label{fig:workflow:rec}
    \end{subfigure}
    \begin{subfigure}{\linewidth}
    \centering
    \includegraphics[width=0.9\linewidth]{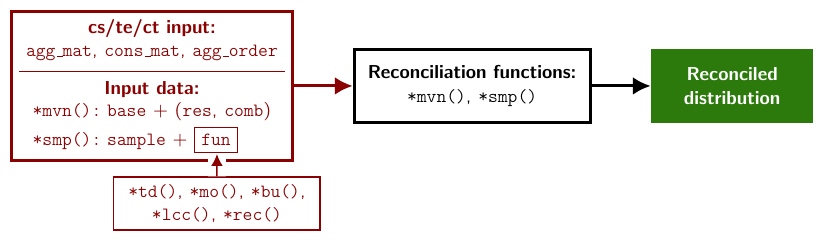}
    \caption{Probabilistic reconciliation workflow} \label{fig:workflow:prob}
    \end{subfigure}
    \begin{subfigure}{\linewidth}
    \centering
    \includegraphics[width=0.9\linewidth]{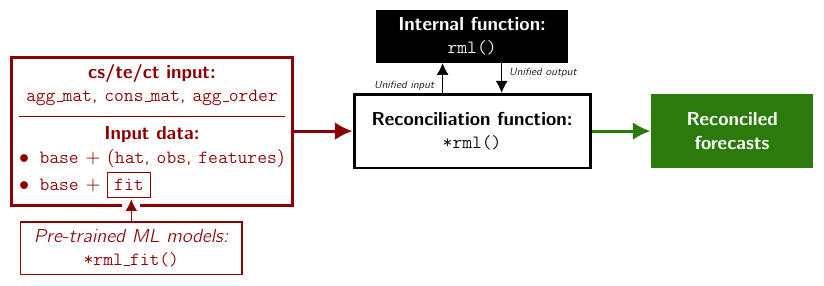}
    \caption{ML-based reconciliation workflow} \label{fig:workflow:rml}
    \end{subfigure}
    \caption{Reconciliation workflows. Panels (\ref{fig:workflow:class}), (\ref{fig:workflow:rec}), and (\ref{fig:workflow:prob}) illustrates the main workflow of \pkg{FoReco}, summarizing the classical, regression-based, and probabilistic reconciliations. Panel (\ref{fig:workflow:rml}) presents the \pkg{FoRecoML} workflow. 
    }
    \label{fig:workflow}
\end{figure}

Figure \ref{fig:workflow} 
illustrates the core workflow common to  \pkg{FoReco} and \pkg{FoRecoML}, thereby emphasizing their shared design philosophy.
The user interface is intentionally simple. 
Each reconciliation function, be it for cross-sectional (\code{cs}), temporal (\code{te}) or cross-temporal (\code{te}) frameworks, 
takes  framework-specific arguments as input: the aggregation matrix (\code{agg_mat}) or constraint matrix (\code{cons_mat}), and the temporal aggregation orders (\code{agg_order}).
{In addition, base forecasts are expected to be generated externally, as the packages are designed to focus on reconciling them to ensure coherency across the system. All functions therefore input the base forecasts via the universal argument \code{base} (a vector, matrix or array, depending on the framework and approach). Beyond \code{base}, each function requires the additional information needed to determine the reconciliation map, and this information differs across approaches. In \pkg{FoReco}, the linear reconciliation functions accept the in-sample residuals through the argument \code{res}, from which the forecast-error covariance matrix is estimated. In \pkg{FoRecoML}, the non-linear reconciliation functions are instead trained on past data and accept the corresponding base forecasts (\code{hat}) and observed values (\code{obs}) used to learn the reconciliation map. These approach-specific inputs are detailed below.}


{This input convention reflects a deliberate design choice: the relevant inputs of all reconciliation functions in \pkg{FoReco} and \pkg{FoRecoML}, including base forecasts, aggregation/constrained matrices and related quantities, are passed as separate standard \proglang{R} objects rather than as a single model or forecast object.}
This keeps the packages agnostic with respect to the upstream forecasting framework: any procedure able to produce point forecasts and the corresponding errors can be used, regardless of the language or library in which the models are fitted. It also accommodates the common practice of estimating the forecast-error covariance matrix from validation errors obtained through a rolling-window scheme \citep[see, e.g.,][]{Abolghasemi2025-fp} rather than from in-sample residuals, a setting in which the base forecasts and the residuals are produced by two distinct procedures and no single object carrying both is available. Since the reconciliation step itself only requires the base forecasts and an estimate of their error covariance, this minimal interface reflects exactly what the method needs.


Given these inputs, the package reconciles the base forecasts and returns an object of class \code{foreco} that contains the reconciled coherent forecasts together with metadata describing the reconciliation procedure (framework, function used, covariance approximation, ML approach, non-negativity setting, and other reconciliation-specific information). To keep the output as close as possible to the most natural representation of the result, the class \code{foreco} extends a numeric matrix or vector for point reconciliation and a distributional object (from the \pkg{distributional} package, \citealp{distributional}) for probabilistic reconciliation. Dedicated \code{print()}, \code{summary()} and \code{plot()} methods are provided for the class, allowing users to inspect the reconciliation output, while the \code{components()} function allows us to split the reconciled forecasts by cross-sectional variables and/or temporal aggregation order. Hence, the overall workflow across the reconciliation functions in \pkg{FoReco} and \pkg{FoRecoML} follows a straightforward ``base forecasts in, reconciled forecasts out'' structure.

Table \ref{tab:ftc-overview} provides an overview of the main functions available in \pkg{FoReco} for linear forecast reconciliation and in \pkg{FoRecoML} for non-linear forecast reconciliation.
The two packages share the same underlying philosophy in the naming convention of their functions: 
The prefix of each function name indicates the reconciliation framework to which it applies -- 
\code{cs*} for cross-sectional,
\code{te*} for temporal, and
\code{ct*} for cross-temporal reconciliation --
while the suffix indicates the specific reconciliation approach used.
Eight reconciliation approaches are implemented: three classical, two regression-based and two probabilistic approaches in \pkg{FoReco}, and non-linear reconciliation in \pkg{FoRecoML}.
The consistent naming structure makes the syntax intuitive and facilitates seamless use across reconciliation frameworks, approaches and the two packages. 

\begin{table}[t]
	\centering
    \resizebox{\linewidth}{!}{
    \begin{tabular}{cm{0\linewidth}m{0.375\linewidth}|>{\centering\arraybackslash}p{7em}>{\centering\arraybackslash}p{7em}>{\centering\arraybackslash}p{7em}}
    \toprule
        & &&\multicolumn{3}{c}{\textbf{Framework}}\\
       \textbf{Package} & \multicolumn{2}{l|}{\textbf{Description}} & Cross-sectional & Temporal & Cross-temporal\\
       \midrule
\pkg{FoReco} & \multicolumn{4}{l}{\textit{Classical reconciliation}}\\
& & Top-down \code{*td()} & \code{cstd()} & \code{tetd()} & \code{cttd()}\\
& & Bottom-up \code{*bu()} & \code{csbu()} & \code{tebu()} & \code{ctbu()}\\
& & Middle-out \code{*mo()} & \code{csmo()} & \code{temo()} & \code{ctmo()}\\
\addlinespace[0.5em]
\pkg{FoReco} & \multicolumn{5}{l}{\textit{Regression-based reconciliation}}\\	
& & Least squares \code{*rec()} & \code{csrec()} & \code{terec()} & \code{ctrec()}\\
& & Level conditional coherent \code{*lcc()} & \code{cslcc()} & \code{telcc()} & \code{ctlcc()}\\
\addlinespace[0.5em]
\pkg{FoReco} & \multicolumn{5}{l}{\textit{Probabilistic reconciliation}}\\
& & Gaussian-based \code{*mvn()} & \code{csmvn()} & \code{temvn()} & \code{ctmvn()}\\
& & Sample-based \code{*smp()} & \code{cssmp()} & \code{tesmp()} & \code{ctsmp()}\\
\addlinespace[0.5em]
\pkg{FoRecoML} & \multicolumn{2}{l|}{\textit{ML-based reconciliation} \code{*rml()}} & \code{csrml()} & \code{terml()} & \code{ctrml()}\\
      \bottomrule
    \end{tabular}}
\caption{\label{tab:ftc-overview}Overview of main reconciliation functions in \pkg{FoReco} and \pkg{FoRecoML}.} 
\vspace{-1em}
\end{table}

In the following, we detail the functions for the different reconciliation approaches across the three reconciliation frameworks. 

\subsection[Classical]{Classical reconciliation in \pkg{FoReco}}  \label{subsec:FoReco-classical}

\pkg{FoReco} offers functionality for all three classical reconciliation approaches: \code{*td()} for top-down, \code{*bu()} for bottom-up and \code{*mo()} for middle-out reconciliation. 

\paragraph{Top-down reconciliation} To perform top-down reconciliation for cross-sectional, temporal or cross-temporal frameworks, \pkg{FoReco} offers respectively: 
\begin{Code}
cstd(base, agg_mat, weights, normalize = TRUE)
tetd(base, agg_order, weights, tew = "sum", normalize = TRUE)
cttd(base, agg_mat, agg_order,  weights, tew = "sum", normalize = TRUE)
\end{Code}

All functions require base forecasts for a given forecast horizon $h$; to be provided via the vector argument \code{base}:  
top-level base forecasts for \code{cs*} frameworks,
temporal aggregated base forecasts of order $m$ for \code{te*} frameworks,
and top- and temporal aggregated level base forecasts
for \code{ct*} frameworks.
After this key argument, framework-related arguments always precede those specific to the reconciliation approach.
The matrix-argument \code{agg_mat} contains the aggregation restrictions encoded in the aggregation matrix $\bm{A}_{cs}$ (see Equation~\ref{eq:cs-A}) and thus needs to be provided for \code{cs*} and \code{ct*} frameworks.
The arguments \code{agg_order} and \code{tew} are specific to \code{te*} and \code{ct*} frameworks: 
 \code{agg_order} supports scalar or vector input since users can choose to either provide the maximal temporal aggregation order $m$, or a vector containing (a subset of) aggregation factors $\{k_p, \ldots, k_1\}$. 
The string argument \code{tew} controls the type of temporal aggregation: \code{"sum"} for summation (default, as in Equation~\ref{eq:ct}), \code{"avg"} for average, \code{"first"} for first value or  \code{"last"} for last value of the period.
The arguments \code{weight} and \code{normalize} are specific to the reconciliation approach:
\code{weights} contains the  proportions for the bottom  series in \code{cs*}, 
for the high-frequency series in \code{te*}, 
and for each high-frequency bottom  series in \code{ct*}; if \code{normalize = TRUE}, then the weights will sum to one.
The functions return, as output, cross-sectional, temporal, or cross-temporal reconciled forecasts.

\paragraph{Bottom-up reconciliation}
For bottom-up reconciliation, \pkg{FoReco} provides:
\begin{Code}
csbu(base, agg_mat, sntz = FALSE, round = FALSE)
tebu(base, agg_order, tew = "sum", sntz = FALSE, round = FALSE)
ctbu(base, agg_mat, agg_order, tew = "sum", sntz = FALSE, round = FALSE)
\end{Code}
In the following, we focus only on the aspects that differ from the \code{*td()} functions discussed above.
For bottom-up reconciliation, the base forecasts for the bottom level for \code{cs*},
the high-frequency series for \code{te*},
and the high-frequency bottom level series for \code{ct*} 
need to be supplied via the argument \code{base}.
The optional logical arguments \code{sntz}  and \code{round} respectively 
let users set negative base forecasts to zero 
or round base forecasts (when \code{TRUE}) 
before applying bottom-up reconciliation.

\paragraph{Middle-out reconciliation}
Middle-out reconciliation is performed using the functions:
\begin{Code}
csmo(base, agg_mat, weights, id_rows = 1, normalize = TRUE)
temo(base, agg_order, weights, order = max(agg_order), tew = "sum", 
     normalize = TRUE)
ctmo(base, agg_mat, agg_order, weights, id_rows = 1, order = max(agg_order), 
     tew = "sum", normalize = TRUE)
\end{Code}
The argument \code{base} now requires base forecasts for a specific level $\ell$ for \code{cs*} reconciliation,
for a specific temporal aggregation order $k$ for \code{te*} reconciliation, and
for level $\ell$ at  temporal aggregation order $k$ for \code{ct*} reconciliation.
The level $\ell$ 
needs to be specified via the vector argument \code{id_rows} which indexes all the rows of the aggregation matrix linked to level $\ell$, whereas  the  temporal aggregation order $k$ via the scalar argument \code{order}.
Since middle-out reconciliation combines a bottom-up approach above level $l/k$ with a top-down approach below, the arguments related to the latter (\code{weights} and \code{normalize}) also need to be supplied.

\subsection[Regression-based reconciliation in FoReco]{Regression-based reconciliation in \pkg{FoReco}}  \label{subsec:FoReco-reg}
\pkg{FoReco} offers two regression-based reconciliation approaches: least-squares based reconciliation via \code{*rec()} and level conditional coherent reconciliation via \code{*lcc()}.

\paragraph{Least-squares based reconciliation}
Least-squares based reconciliation can be performed using the functions \code{csrec()}, \code{terec()}, and \code{ctrec()}.
We introduce function \code{ctrec()} in full detail, as it covers the complete set of arguments for cross-temporal reconciliation. Functions \code{csrec()} and \code{terec()} are special cases, restricted to cross-sectional and temporal frameworks, respectively; at the end, we therefore briefly highlight the arguments relevant for each of these cases. 
\begin{Code}
ctrec(base, agg_mat, cons_mat, agg_order, tew = "sum", comb = "ols", 
      res = NULL, approach = "proj", nn = NULL, settings = NULL, 
      bounds = NULL, immutable = NULL, ...)       
\end{Code}
The argument \code{base} now requires users to specify the base forecasts $\widehat{\bf{X}}$ at any temporal frequency for all the series;
the forecasts for cross-sectional units are arranged in the rows, and those for different levels of temporal aggregation in the columns.
The cross-sectional aggregation constraints can either be specified using the 
 the aggregation matrix $\bm{A}_{cs}$ (see Equation~\ref{eq:cs-A}) or
the constraint matrix $\bm{C}_{cs}$ (see Equation~\ref{eq:cs-C}) via the argument \code{cons_mat}.
These arguments also need to be supplied in the function \code{csrec()}.
The temporal aggregation constraints, controlled via \code{agg_order} and \code{tew}, work as described above and must also be supplied to the \code{terec()} function.

The remaining arguments are specific to least-squares based reconciliation and hence need to be provided for all reconciliation frameworks.
The string argument \code{comb} controls the choice of base forecast covariance matrix; Table~\ref{tab:Omega} gives the available options for each across the reconciliation frameworks. 
The default is \code{"ols"}, selected for its simplicity and because it avoids the need for a user-supplied residual matrix via the argument \code{res}.
Weighted least squares can be used either with or without user-supplied residual matrix, whereas
generalized least squares options always requires the user to specify \code{res}; see Table~\ref{tab:Omega}  for an overview.

The string argument \code{approach} lets users choose between the projection approach or the structural approach (default \code{"proj"} versus \code{"strc"}; see Section \ref{subsec:methods}). 
 The two approaches are theoretically equivalent but differ in computational efficiency, with the projection method typically offering faster performance \citep{nn2025}; hence, it is set as the default option.
\pkg{FoReco} also supports numerical solution for both using the operator splitting quadratic program (OSQP; \citealp{osqp}) solver (\code{"proj_osqp"} or \code{"strc_osqp"} respectively). These solver-based options are particularly useful when analytical solutions are not straightforward or when enforcing non-linear constraints (e.g., non negativity), allowing users to directly solve the constrained optimization problem.

The additional arguments all support further extensions to compute non-negative forecasts (string argument \code{nn}), to impose lower and/or upper limits on the reconciled forecasts (matrix argument \code{bounds}), and to ensure certain base forecasts remain fixed, or immutable (matrix argument \code{immutable}; \citealp{Zhang2023-yj}). Specifically, the string argument \code{nn} allows users to select among five algorithms for non-negative reconciliation: quadratic programming via the OSQP solver (\code{"osqp"}; \citealp{osqp, nn2025}), the block principal pivoting algorithm (\code{"bpv"}; \citealp{Wickramasuriya2020-zk}), a negative forecasts correction algorithm (\code{"nfca"}; \citealp{Kourentzes2021-lx, nn2025}), an iterative method with immutable constraints (\code{"nnic"}; \citealp{nn2025}), and a heuristic set-negative-to-zero approach (\code{"sntz"}; \citealp{Di_Fonzo2023-ae, nn2025}). Additional control parameters for each algorithm can be specified through \code{settings}. 

Finally, although least-squares-based reconciliation is performed automatically by the main functions \code{csrec()}, \code{terec()}, and \code{ctrec()} that all return reconciled forecasts, \pkg{FoReco} also provides functions to approximate the base forecast covariance matrix -- namely, \code{cscov()}, \code{tecov()}, and \code{ctcov()}, whose output is a symmetric positive (semi-)definite covariance matrix.
The main functions internally call these routines to approximate the base forecast covariance matrix, but the stand-alone functions may also be of independent interest to expert users. The  argument \code{...} in the main functions \code{csrec()}, \code{terec()}, and \code{ctrec()} allows users to further customize their own covariance matrix if desired.

For cross-temporal reconciliation, the covariance function is given by 
\begin{Code}
ctcov(comb = "ols", agg_mat = NULL, agg_order = NULL, tew = "sum", 
      res = NULL, n = NULL, mse = TRUE, shrink_fun = shrink_estim, ...)
\end{Code}
The arguments listed on the first line also appear in the main function and behave in the same way, with \code{agg_mat} also needed for \code{cscov()} and \code{agg_order} and \code{tew}  needed for \code{tecov()}.
The arguments listed on the second line appear for all reconciliation frameworks. 
The scalar argument \code{n} sets the number of cross-sectional variables.
The logical argument \code{mse} is only relevant when \code{res} is provided; when set to \code{TRUE} the residuals are not mean-corrected.
The argument \code{shrink_fun} allows users to provide a custom shrinkage function for the covariance matrix; it takes the residual matrix as input and returns the covariance matrix. The argument \code{...} in the function enables users to pass additional parameters, allowing them to specify or customize their own shrinkage function as needed.

In addition to LS–based approaches, \pkg{FoReco} also provides heuristic and iterative alternatives \citep{Kourentzes2019-dj,Di_Fonzo2023-dg}  for cross-temporal reconciliation, implemented through the functions \code{tcsrec()}, \code{cstrec()}, and \code{iterec()}:
\begin{Code}
tcsrec(base, cslist, telist, res = NULL, avg = “KA”)
cstrec(base, cslist, telist, res = NULL)
iterec(base, cslist, telist, res = NULL, itmax = 100, tol = 1e-5,
       type = “tcs”, norm = “inf”, verbose = TRUE)
\end{Code}
These functions are not included in Table \ref{tab:ftc-overview}, which summarizes the main reconciliation functions, as they are heuristic in nature—though still valuable in practice.
 The two-step heuristic functions \code{tcsrec()} and \code{cstrec()} perform reconciliation in a fixed order, either first temporal and then cross-sectional (\code{tcsrec()}) or vice versa (\code{cstrec()}). The iterative \code{iterec()} performs reconciliation alternately along one dimension at a time until convergence, allowing repeated adjustment between temporal and cross-sectional coherence. All require  \code{base} as input, as provided by \code{ctrec()}. The lists \code{cslist} and \code{telist} collect the arguments for the cross-sectional and temporal reconciliation steps, respectively, corresponding to those of \code{csrec()} and \code{terec()} (except \code{base} and \code{res}). The optional argument \code{res} allows users to provide in-sample residuals or validation errors. In \code{tcsrec()}, the argument \code{avg = “KA”} (default) applies the KA projection averaging; alternatively, a simple average of the projection matrices can be used. In \code{iterec()} additional arguments can be specified: \code{itmax}, which sets the maximum number of iterations; \code{tol}, defining the convergence tolerance; \code{type}, specifying the iteration order (\code{“tcs”} or \code{“cst”}); and \code{norm}, which selects the incoherence norm used to check convergence (\code{“inf”} by default, with \code{“one”} or \code{“two”} also available); the argument \code{verbose} controls whether iteration progress is printed. 

\paragraph{Level conditional coherent reconciliation}
Level conditional coherent reconciliation can be performed using the functions \code{cslss()}, \code{telcc()} and \code{ctlss()}. 
For brevity, we again discuss the function for cross-temporal reconciliation in detail: 
\begin{Code}
ctlcc(base, agg_mat, agg_order, tew = "sum", comb = "ols", 
      res = NULL, approach = "proj", nn = NULL, settings = NULL, 
      CCC = TRUE, const = "exogenous", hfbts = NULL, ...)
\end{Code}
All arguments on the first two lines play the same role as in the \code{ctrec()} function, the remaining ones are specific to level conditional coherent reconciliation.
The logical argument \code{CCC} specifies whether the combined conditional coherent reconciled forecasts reconciliation should include bottom-up forecasts (\code{TRUE}) or not. 
The string argument \code{const} can take on two values: \code{"exogenous"} to fix the top level of each sub-hierarchy (default) or \code{"endogenous"} to coherently revise the top and bottom levels.
Lastly, users can optionally provide alternative base forecasts for the lowest-level series through the argument \code{hfbts} (or \code{bts} and \code{hfts} in the cross-sectional and temporal frameworks, respectively), following the approach of \cite{Hollyman2021-zq}. If omitted, the function uses only the base forecasts supplied in \code{base}. 

\subsection[Probabilistic reconciliation in FoReco]{Probabilistic reconciliation in \pkg{FoReco}}  \label{subsec:FoReco-prob}
\pkg{FoReco} supports two probabilistic reconciliation approaches: a Gaussian-based approach and a sample-based approach.

\paragraph{Gaussian-based probabilistic reconciliation} To perform probabilistic forecast reconciliation assuming a multivariate normal base forecast distribution, the functions \code{csmvn()}, \code{temvn()} and \code{ctmvn()} can be used. The latter is structured as follows (\code{csmvn()} and \code{temvn()} are structured similarly):
\begin{Code}
ctmvn(base, agg_mat, cons_mat, agg_order, tew = "sum", comb = "ols", 
      res = NULL, approach = "proj", comb_base = comb, 
      reduce_form = FALSE, ...)
\end{Code}
All arguments, apart from \code{comb_base} and \code{reduce_form}, are consistent with those of the \code{ctrec()} function, which \code{ctmvn()} extends in a natural probabilistic manner.
By default, the type of covariance matrix for the base forecasts is the same as that used for all series (\code{comb_base = comb}; $\bm\Sigma = \bm\Omega$, see Section \ref{subsec:methods}). However, users may specify a different type of covariance matrix, as the argument \code{comb_base} accepts the same options listed in Table~\ref{tab:Omega}.
The logical argument \code{reduce_form} determines whether the full distribution (\code{FALSE}, default) or only that of the high-frequency bottom time series is returned (bottom-level for \code{csmvn()}, high-frequency for \code{temvn()}).
The function returns an object of class \code{foreco} whose underlying distributional object is of class \code{dist_multivariate_normal} from the \pkg{distributional} package \citep{distributional}, allowing a more consistent and flexible representation of multivariate distributions than a simple matrix.

\paragraph{Sample-based probabilistic reconciliation} The functions \code{cssmp()}, \code{tesmp()} and \code{ctsmp()} offer functionality to perform probabilistic reconciliation using a sample based approach. Their structure is simple:
\begin{Code}
cssmp(sample, fun = csrec, ...)
tesmp(sample, agg_order, fun = terec, ...)
ctsmp(sample, agg_order, fun = ctrec, ...)
\end{Code}
Given base forecast samples specified via \code{sample}  
the functions 
apply a chosen \pkg{FoReco} reconciliation approach independently to each draw, thereby producing a coherent sample distribution of reconciled forecasts across the cross-sectional, temporal or cross-temporal framework.
The default is to use  least-squares based reconciliation. 
The function returns an object of class \code{foreco} whose underlying distributional object is of class \code{dist_sample} from the \pkg{distributional} package. 

\subsection[ML-based reconciliation in FoRecoML]{ML-based reconciliation in \pkg{FoRecoML}} \label{subsec:FoRecoML}

\pkg{FoRecoML} offers nonlinear forecast reconciliation using machine learning for cross-sectional, temporal, and cross-temporal frameworks using the functions \code{csrml()}, \code{terml} and \code{ctrml()} which all return reconciled forecasts. 

For cross-sectional frameworks, the function is structured as
\begin{Code}
csrml(base, hat, obs, agg_mat, features = "all", approach = "randomForest", 
      params = NULL, tuning = NULL, sntz = FALSE, round = FALSE, fit = NULL)
\end{Code}
The matrix argument \code{base} contains the base forecasts that need to be reconciled, and \code{agg_mat} the aggregation matrix $\bm{A}_{cs}$, as before. If the cross-sectional constraints are encoded in the constraint matrix $\bm{C}_{cs}$ instead, the function \code{lcmat()} in \pkg{FoReco} can be used to obtain the aggregation matrix through a suitable transformation of $\bm{A}_{cs}$. 
The additional matrix argument \code{hat} and \code{obs}, and the string argument \code{features} are specific to the validation sample of the ML-based approach: \code{hat} contains the base forecasts of all series on the validation sample whereas \code{obs} contains the  actual (observed) values of the bottom-level series on the same validation sample; the number of rows (i.e., time points in the validation sample) thus needs to be the same in both. 
The string argument \code{features} specifies which features are used for model training and defaults to using all series (\code{"all"}). Other options are to use only the bottom-level series (\code{"bts"}), the series based on the structural matrix (\code{"str"}); namely all series on the path from a bottom-level series to the top-level series (e.g., toy example of Figure~\ref{fig:toy:cs}: to train a model for $y_4$, the series $y_1, y_2, y_4$ are used as features), or the combination of the latter two (\code{"str-bts"}; e.g., toy example of Figure~\ref{fig:toy:cs}: to train a model for $y_4$, the series $y_1, y_2$ and $y_4$ to $y_8$ are used as features).

Next, the string argument \code{approach} specifies the ML method to be used. Supported options are \code{"randomForest"} from the \pkg{randomForest} package,
\code{"xgboost"} from the \pkg{xgboost} package, \code{"lightgbm"} from the \pkg{lightgbm} package and \code{"mlr3"} which allows users to use any regression learner available in the \pkg{mlr3} package.
The optional list arguments \code{params} allow users to pass on additional parameters to the chosen ML approach such as algorithm-specific hyperparameters, the list argument \code{tuning} is only relevant when \code{approach = "mlr3"} as it allows users to specify further tuning options.
The logical options \code{sntz} (the "set-negative-to-zero" heuristic) and \code{round} (rounding reconciled forecasts to integers) work in the same way as in the \code{*bu} functions used for bottom-up reconciliation, since the ML-based approach relies on a bottom-up reconciliation to produce coherent forecasts.

Lastly, the \code{fit} argument in \code{csrml()} 
allows users to pass on a pre-trained ML reconciliation model. If supplied, this  omits the need for providing \code{hat} and \code{obs}.  
To obtain a pre-trained ML model, one can use the function \code{csrml_fit}, with the same arguments as \code{csrml()}, namely
\begin{Code}
csrml_fit(hat, obs, agg_mat, features = "all", approach = "randomForest", 
           params = NULL, tuning = NULL)
\end{Code}
The function returns a list that contains the trained ML model. The latter can then, in turn, be re-used for reconciliation on new base forecasts via the argument \code{fit} in \code{csrml()}.

The functions to perform ML-based temporal  or cross-temporal reconciliation work very similarly and are given by 
\begin{Code}
terml(base, hat, obs, agg_order, tew = "sum", features = "all", 
      approach = "randomForest", params = NULL, tuning = NULL, sntz = FALSE, 
      round = FALSE, fit = NULL)
ctrml(base, hat, obs, agg_mat, agg_order, tew = "sum", features = "all",
      approach = "randomForest", params = NULL, tuning = NULL, sntz = FALSE, 
      round = FALSE, fit = NULL)      
\end{Code}
The arguments \code{agg_order} and \code{tew} specify the temporal aggregation constraints, as in the linear reconciliation functions. All other arguments function similarly to those in \code{csrml()}. 
For temporal frameworks, the \code{features} argument can be set to either use all series (\code{"all"}, the default) or to use only the lowest- and highest-frequency base forecasts (\code{"low-high"}).
For cross-temporal frameworks, the \code{features} argument can similarly be set to \code{"all"} (default) or to \code{"compact"}, the latter uses a reduced feature matrix in which, for the temporal dimension, only the distinct temporal frequencies of the bottom-level series used to train the model are retained; see \cite{Rombouts2025-ym}.

\section{Illustration with data}\label{sec:illustration-with-data}

We illustrate the usage of \pkg{FoReco} and \pkg{FoRecoML} with real
data in three different settings: cross-sectional, temporal and
cross-temporal forecast reconciliation. We work through examples using
the Australian tourism data in the cross-sectional setting, and Italian
energy load data in the temporal and the cross-temporal setting, showing
how to obtain base forecasts and reconcile these forecasts with
different reconciliation approaches.

\subsection{Cross-sectional forecast
reconciliation}\label{cross-sectional-forecast-reconciliation}

\paragraph{Australian tourism}

We first load the \pkg{FoReco} package for reconciliation and the
\pkg{forecast} package \citep{Hyndman-2008} to generate base forecasts.

\begin{CodeChunk}

\begin{CodeInput}
R> library("FoReco")
R> library("forecast")
\end{CodeInput}

\end{CodeChunk}

We use the \code{vndata} dataset, which contains Australian tourism
demand dataset, and \code{vnaggmat}, which is the corresponding
aggregation matrix. The Australian Tourism Demand dataset
\citep{Wickramasuriya2019, Girolimetto2024-jm} contains the number of
nights Australians spent away from home. It includes 228 monthly
observations of Visitor Nights (VNs) from January 1998 to December 2016,
and has a cross-sectional grouped structure based on a geographic
hierarchy crossed by \(4\) purposes of travel.
Table~\ref{tbl-tourism-stru} shows the number of series across each
geographic region and travel purpose category; in total we have
\(n=525\) series.

\begin{table}

\centering{

\begin{tabular}[t]{lrrrrr}
\toprule
  & AUS & States & Zones & Regions & Total\\
\midrule
Geographical divisions & 1 & 7 & 21 & 76 & 105\\
Purpose of travel & 4 & 28 & 84 & 304 & 420\\
Total & 5 & 35 & 105 & 380 & 525\\
\bottomrule
\end{tabular}

}

\caption{\label{tbl-tourism-stru}Australian tourism demand: Number of
series by geographic region (columns) and travel purpose (rows).}

\end{table}%

We generate base forecasts using ETS models. A log transformation on
each series is applied to ensure positive forecasts. Zero values are
replaced by half the minimum non-zero value in the series
\citep{Wickramasuriya2020-zk}.

\begin{CodeChunk}

\begin{CodeInput}
R> model <- fc_obj <- vector("list", length = NCOL(vndata))
R> names(model) <- names(fc_obj) <- colnames(vndata)
R> ets_log <- function(x, ...) {
+    x[x == 0] <- min(x[x != 0]) / 2
+    ets(x, lambda = 0, ...)
+  }
R> for (i in 1:NCOL(vndata)) {
+    model[[i]] <- ets_log(vndata[, i])
+    fc_obj[[i]] <- forecast(model[[i]], h = 12)
+  }
\end{CodeInput}

\end{CodeChunk}

We extract point forecasts \code{base} and residuals \code{res} from the
fitted models for use in the reconciliation approaches.

\begin{CodeChunk}

\begin{CodeInput}
R> base <- do.call(cbind, lapply(fc_obj, function(x) x$mean))
R> res <- do.call(cbind, lapply(fc_obj, residuals, type = "response"))
\end{CodeInput}

\end{CodeChunk}

\paragraph{Linear reconciliation}

We apply various reconciliation approaches to the base forecasts.
Bottom-up reconciliation aggregates forecasts from the lowest level to
higher levels. The \code{print.foreco()} method allows users to specify
the number of rows and columns of reconciled forecasts to examine. The
\code{summary.foreco()} method also reports the reconciliation framework
(cross-sectional, temporal or cross-temporal), the function that
produced the object, the forecast type (point or probabilistic), the
covariance approximation \code{comb}, the machine-learning approach
\code{ml} (when applicable), and the non-negativity setting \code{nn};
when a non-negative reconciliation algorithm has been applied, the
convergence status returned by the algorithm is reported as well.

\begin{CodeChunk}

\begin{CodeInput}
R> fc_bts <- base[, colnames(vnaggmat)]
R> rf_bu <- csbu(fc_bts, agg_mat = vnaggmat)
R> summary(rf_bu)
\end{CodeInput}

\begin{CodeOutput}
v Cross-sectional point forecast reconciliation

-- Method 
* Function used: `csbu`
* Output: (12 x 525) matrix

-- Structure 
* Number of cross-sectional series: 525
* Forecast horizons (h): 12
* Non-negative forecasts (check): `TRUE`

-- Reconciled forecasts
       Total         A         B        C        D        E
h-1 44094.08 15090.110 10589.205 8964.637 2855.082 4821.167
h-2 18218.45  5881.055  3793.292 3994.225 1106.625 2516.225
h-3 20585.12  6592.540  4482.910 4183.151 1301.891 2849.777
h-4 25563.48  8609.232  5174.437 5631.049 1729.393 3352.184
... (8 more rows, 519 more columns)
Use `print(x, n_row, n_col)` to see more rows and columns.
\end{CodeOutput}

\end{CodeChunk}

\begin{figure}[t]

\centering{

\includegraphics[width=1\linewidth,height=\textheight,keepaspectratio]{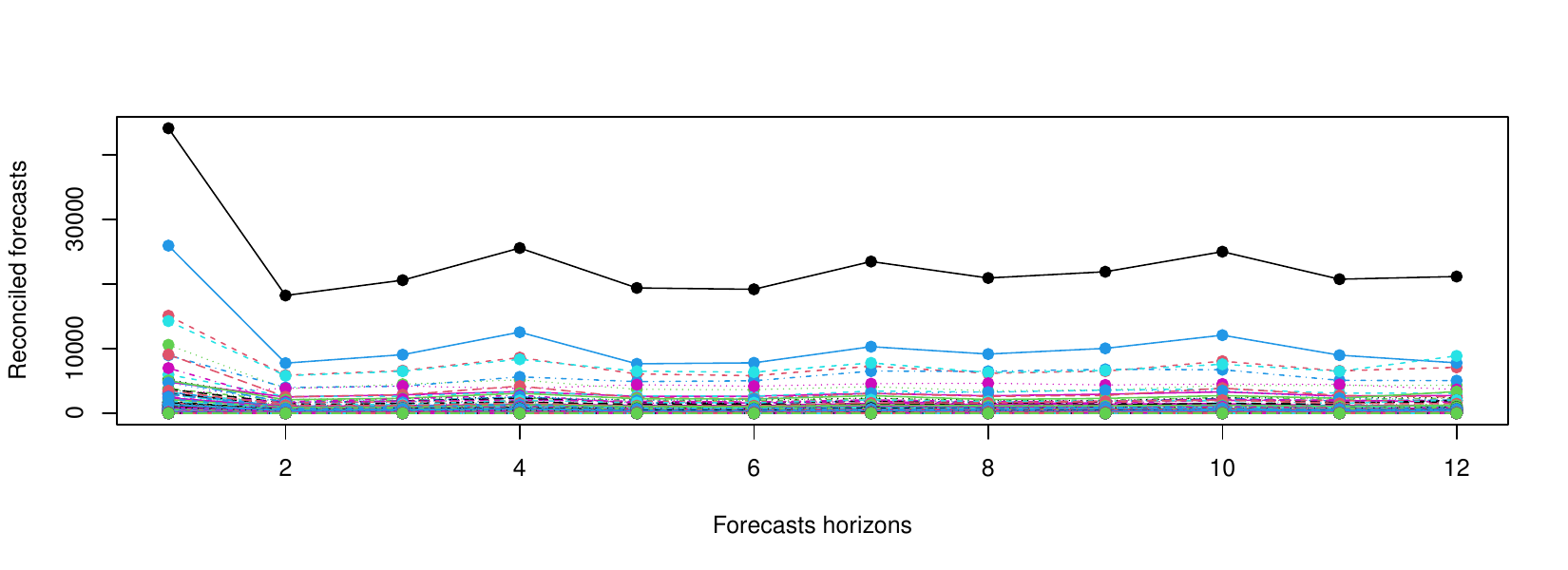}

}

\caption{\label{fig-plot}Bottom-up reconciled forecast plotted using the
\code{plot.foreco()} method.}

\end{figure}

The \code{plot.foreco()} method offers a graphical inspection of the
reconciled output (Figure~\ref{fig-plot}).

\begin{CodeChunk}

\begin{CodeInput}
R> plot(rf_bu)
\end{CodeInput}

\end{CodeChunk}

Top-down reconciliation disaggregates according to a proportional
scheme. We use the average historical proportions as the disaggregation
weights.

\begin{CodeChunk}

\begin{CodeInput}
R> bts <- vndata[, colnames(vnaggmat)]
R> total <- vndata[, "Total"]
R> fc_total <- base[, "Total"]
R> p_gsa <- colMeans(apply(bts, 2, function(x) x / total))
R> rf_td_gsa <- cstd(fc_total, agg_mat = vnaggmat, weights = p_gsa)
R> print(rf_td_gsa, n_row = 2, n_col = 6)
\end{CodeInput}

\begin{CodeOutput}
       Total         A        B         C        D        E
h-1 50651.48 16251.582 9659.413 13361.324 3417.451 5239.247
h-2 21335.64  6845.562 4068.780  5628.115 1439.514 2206.899
... (10 more rows, 519 more columns)
Use `print(rf_td_gsa, n_row, n_col)` to see more rows and columns.
\end{CodeOutput}

\end{CodeChunk}

For LCC reconciliation, the reconciled forecasts are constrained by the
base forecasts of a specific upper level of the hierarchy. In
\pkg{FoReco}, the default is to fix the top level of each sub-hierarchy.

\begin{CodeChunk}

\begin{CodeInput}
R> rf_lcc <- cslcc(base = base, agg_mat = vnaggmat, res = res, comb = "wls")
R> print(rf_lcc, n_row = 2, n_col = 6)
\end{CodeInput}

\begin{CodeOutput}
       Total         A         B        C        D        E
h-1 47896.34 16171.607 11489.265 9986.096 3107.838 5106.384
h-2 20404.61  6464.813  4265.185 4642.466 1270.962 2668.350
... (10 more rows, 519 more columns)
Use `print(rf_lcc, n_row, n_col)` to see more rows and columns.
\end{CodeOutput}

\end{CodeChunk}

Finally, we obtain LS-based cross-sectional reconciled forecasts using
two different options for forecast error covariance estimation (see
Table~\ref{tab:Omega}): OLS (i.e., the default for argument
\code{comb}), and shrinkage (i.e., \code{comb = "shr"}).

\begin{CodeChunk}

\begin{CodeInput}
R> rf_ols <- csrec(base = base, agg_mat = vnaggmat)
R> rf_shr <- csrec(base = base, agg_mat = vnaggmat, res = res, comb = "shr")
\end{CodeInput}

\end{CodeChunk}

\subsection{Temporal forecast
reconciliation}\label{temporal-forecast-reconciliation}

\paragraph{Italian energy load data}

To demostrate the use of \pkg{FoReco} and \pkg{FoRecoML} in the temporal
and cross-temporal setting, we use the Italian energy load data. It
provides electricity consumption data for seven bidding zones (Calabria,
Center-North, Center-South, North, Sardinia, Sicily, and South) of Italy
from January, 1 2024 to June, 30 2025, measured in megawatts (MW). We
aggregated the original data (recorded at 15-minute intervals) to 1-hour
intervals for illustration purpose. The temporal hierarchy consists of
\(p=8\) levels with aggregation levels
\(\mathcal{K} = \{1, 2, 3, 4, 6, 8, 12, 24\}\). For temporal forecast
reconciliation, we take data from June, 28 2024 to June, 28 2025 as the
training set, and forecast and reconcile the last \(2\) days in the
dataset.

To generate the base forecasts, we specify a seasonal ARIMA model for
each level. For series at the daily level (\(k=24\)), a weekly
seasonality with a frequency of \(7\) is specified. For series in the
subday level, a seasonality of \(24/k\) is specified. For example, a
series at the two-hour level (\(k=2\)) has a daily seasonality with a
frequency of 12. Six dummy variables representing Tuesday to Sunday are
also included as external regressors, where Monday is taken as the base
category, to accommodate weekly seasonality. An ARIMA model is fitted at
each level using the last \(2000\) observations of the total of Italy,
where the orders of AR and MA terms are selected using AICc with the
\code{auto.arima()} function. These same orders are then applied to all
other series at the corresponding levels, with parameters re-estimated
on their respective training sets.

To approximate the base forecast error covariance matrix, we compute
out-of-sample validation errors \citep{Abolghasemi2025-fp} rather than
in-sample residuals as in
Section~\ref{cross-sectional-forecast-reconciliation}. To calculate
validation errors, we use a rolling window approach. More specifically,
a \(366\)-day training period is used for generating base forecasts,
followed by a \(24\) hours validation period for calculating forecast
errors. Each iteration produces forecasts for the next \(1\) day, with
the rolling window advancing one day at a time. This generates \(179\)
rolling windows for the validation errors. For the \(180\)-th window, we
generated forecasts for the next \(2\) days on which reconciliation is
implemented.

\paragraph{Regression-based reconciliation}

For the temporal forecast reconciliation, we only consider the series of
the total energy load of Italy. The base forecasts are stored in the
vector \code{bf_italy}, arranged as described in
Section~\ref{subsec:frameworks}: \(\left[
x_1^{[k_p]} \;
\dots \;
x_1^{[k]} \; \dots \; x_{m/k}^{[k]} \;
\dots \;
x_1^{[1]} \; \dots \; x_{m}^{1}
\right]^\top\).

The validation errors are stored in \code{err_italy}.

We first demonstrate the regression-based reconciliation approach with
three different forecast error covariance estimation, OLS (\code{ols},
the default), shrinkage (\code{shr}), and weighted least squares
reconciliation with series variances (\code{wlsv}).

\begin{CodeChunk}

\begin{CodeInput}
R> reco_teols <- terec(base = bf_italy, agg_order = 24)
R> reco_teshr <- terec(base = bf_italy, res = err_italy,
+                      agg_order = 24, comb = "shr")
R> reco_tewlsv <- terec(base = bf_italy, res = err_italy,
+                       agg_order = 24, comb = "wlsv")
\end{CodeInput}

\end{CodeChunk}

The output of \code{terec()} is an object of class \code{foreco} that
stores the reconciled forecasts as a single stacked vector, in which the
forecasts of all temporal aggregation orders are concatenated according
to the standard layout adopted in the forecast reconciliation literature
(see Section~\ref{subsec:frameworks}). The \code{components.foreco()}
method can be used to extract forecasts for specific temporal
aggregation orders, where each order \(k \in \mathcal{K}\) corresponds
to \(hm/k\) forecasts. The \(h\) represents the forecast horizon at the
most aggregated level. The output is a named list of reconciled
forecasts split by temporal aggregation order. When only one temporal
order is specified, the argument \code{simplify = TRUE} can be used to
simplify the output to a vector. For the shrinkage-based reconciliation,
to extract the forecasts of order \(24\), we have:

\begin{CodeChunk}

\begin{CodeInput}
R> components(reco_teshr, te = 24, simplify = TRUE)
\end{CodeInput}

\begin{CodeOutput}
[1] 3397223 4559157
\end{CodeOutput}

\end{CodeChunk}

\paragraph{Probabilistic reconciliation}

Probabilistic reconciliation extends the point forecast reconciliation
approaches to full predictive distributions (see
Section~\ref{subsec:methods}). We first perform the Gaussian-based
approach, implemented via the \code{temvn()} function.

\begin{CodeChunk}

\begin{CodeInput}
R> reco_dist <- temvn(base = bf_italy, agg_order = 24, comb = "shr", 
+                     res = err_italy)
\end{CodeInput}

\end{CodeChunk}

\begin{figure}[t]

\centering{

\includegraphics[width=1\linewidth,height=\textheight,keepaspectratio]{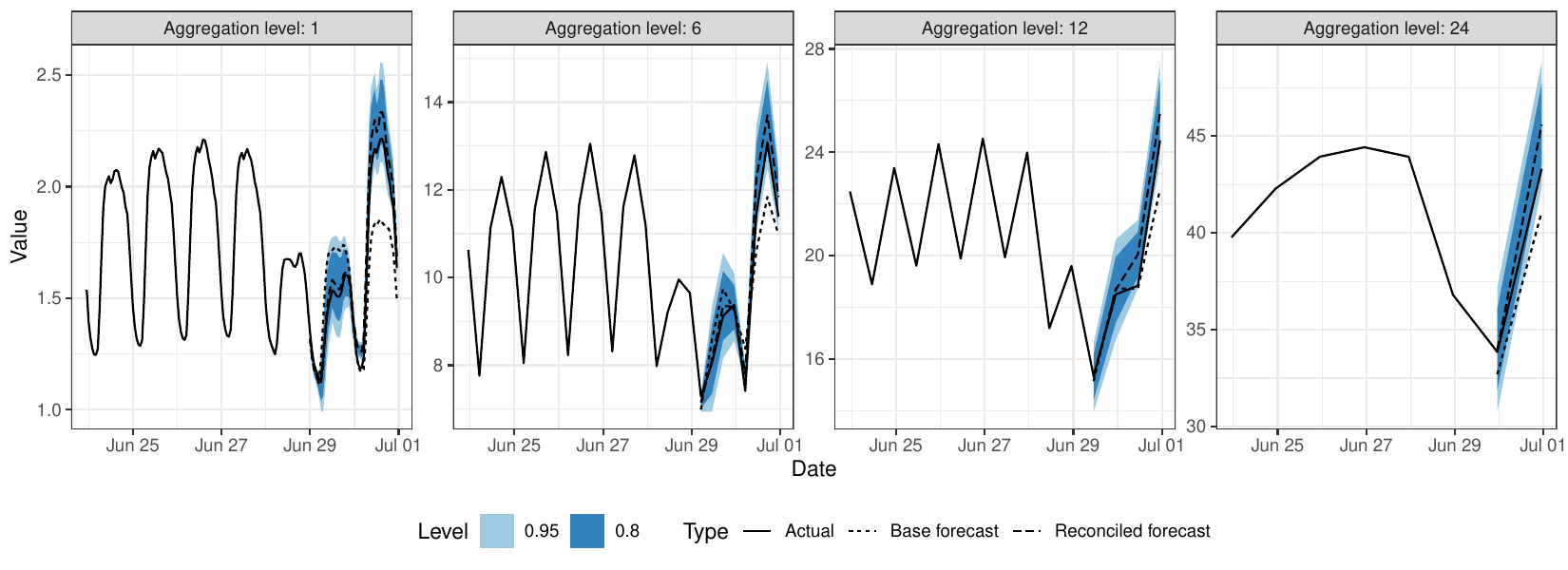}

}

\caption{\label{fig-prob-dist-plot}Italian energy load
(\(\times 10^5\)): Base forecasts versus reconciled probabilistic
forecasts from the Gaussian-based approach.}

\end{figure}

The output is an object of class \code{foreco}, which includes a vector
of distributions. Figure~\ref{fig-prob-dist-plot} compares the base
forecasts with its probabilistic forecasts.

To perform sample-based probabilistic reconciliation, we first bootstrap
possible base forecasts from the base forecast models using the
\code{teboot()} function. \code{model} in the \code{model_list} argument
is a list of base forecasts models ordered from the lowest frequency
(most temporally aggregated) to the highest frequency. We set
\code{block_size = 2} for the forecast horizon of \(2\) days. Dummy
variables are provided in the \code{weekday_dummy} matrix, where each
column represents a dummy variable and rows correspond to levels from
lowest to highest frequency, as described in
Section~\ref{subsec:frameworks}. Variables not used in a level's model
are set to \code{NA}. In our example, the day-level model does not use
any dummy variables, so the first two rows contain \code{NA}s.

\begin{CodeChunk}

\begin{CodeInput}
R> base_boot <- teboot(model, boot_size = 100, agg_order = 24, 
+                      block_size = 2, xreg = weekday_dummy)
\end{CodeInput}

\end{CodeChunk}

Using the bootstrapped base forecast \code{base_boot}, we can then
perform sample-based reconciliation with \code{tesmp()}.

\begin{CodeChunk}

\begin{CodeInput}
R> reco_samp <- tesmp(base_boot, agg_order = 24, res = err_italy, 
+                     comb = "shr")
\end{CodeInput}

\end{CodeChunk}

\subsection{Cross-temporal forecast
reconciliation}\label{cross-temporal-forecast-reconciliation}

We use the Italian energy load dataset with all seven bidding zones and
their aggregate to perform cross-temporal reconciliation. The base
forecasts are generated as described in
Section~\ref{temporal-forecast-reconciliation} and are stored in the
matrix \code{bf_mat}. Each row of \code{bf_mat} corresponds to a
different cross-sectional hierarchy. Here we have \(7\) bidding zones
and the total of Italy, so the row number is \(n = 8\). For this simple
cross-sectional hierarchy, the aggregation matrix \code{agg_mat} is just
a single row of \(7\) ones. The entries in each row are the forecasts of
different temporal hierarchies, arranged the same way as described in
Section~\ref{subsec:frameworks}. The validation errors are arranged
similarly in the matrix \code{val_err_mat}.

\begin{CodeChunk}

\begin{CodeInput}
R> reco_lr <- ctrec(base = bf_mat, res = val_err_mat, agg_order = 24,
+                   agg_mat = matrix(1L, ncol = 7), comb = "shr")
R> print(reco_lr, n_row = 2, n_col = 5)
\end{CodeInput}

\begin{CodeOutput}
           k-24 h-1   k-24 h-2   k-12 h-1   k-12 h-2   k-12 h-3
Italy    3415859.62 4431345.34 1543358.68 1872500.93 1952815.16
Calabria   62894.25   74070.24   26119.56   36774.69   31096.69
... (6 more rows, 115 more columns)
Use `print(reco_lr, n_row, n_col)` to see more rows and columns.
\end{CodeOutput}

\end{CodeChunk}

The specification for regression-based reconciliation approach in the
cross-temporal setting is similar to other settings. The default
projection approach (\code{approach = "proj"}) is, generally, faster
because the matrix multiplication involves smaller matrices \citep[see][
for a comparison of computational times]{nn2025}.

\paragraph{ML-based reconciliation}

We first load \pkg{FoRecoML} for ML-based reconciliation.

\begin{CodeChunk}

\begin{CodeInput}
R> library("FoRecoML")
\end{CodeInput}

\end{CodeChunk}

The ML model used for reconciliation can be trained together with the
reconciliation step. We use extreme gradient boosting from the
\pkg{xgboost} package for illustration purpose. ML-based reconciliation
then requires two additional arguments: \code{val_bf_mat}, which stores
the base forecasts for all series on the validation set, and
\code{val_hfbts}, which contains the observed values of the bottom-level
series at the highest frequency:

\begin{CodeChunk}

\begin{CodeInput}
R> reco_ml <- ctrml(base = bf_mat, hat = val_bf_mat, obs = val_hfbts, 
+                   agg_mat = matrix(1L, ncol = 7), agg_order = 24,
+                   approach = "xgboost")
\end{CodeInput}

\end{CodeChunk}

The ML model can then be extracted from the reconciliation output:

\begin{CodeChunk}

\begin{CodeInput}
R> rf_mdl <- extract_reconciled_ml(reco_ml)
\end{CodeInput}

\end{CodeChunk}

Alternatively, a ML model can be explicitly trained separately:

\begin{CodeChunk}

\begin{CodeInput}
R> rf_mdl <- ctrml_fit(hat = val_bf_mat, obs = val_hfbts, agg_order = 24,
+                      agg_mat = matrix(1L, ncol = 7), approach = "xgboost")
\end{CodeInput}

\end{CodeChunk}

The model can then be used in a subsequent reconciliation step. Instead
of supplying the based forecasts on the validation set and the
highest-frequency bottom-level actual observations, we provide the
pretrained ML model:

\begin{CodeChunk}

\begin{CodeInput}
R> reco_ml <- ctrml(base = bf_mat, fit = rf_mdl, agg_order = 24,
+                   agg_mat = matrix(1L, ncol = 7))
\end{CodeInput}

\end{CodeChunk}

\pkg{FoRecoML} supports \pkg{mlr3} framework, providing access to its
wide range of models. The ML model used for reconciliation can be
specified using the \code{approach} argument, and addtional parameters
can be passed to the ML model through the \code{params} argument. We
illustrate this with extreme gradient boosting
(\code{.key = "regr.xgboost"}).

\begin{CodeChunk}

\begin{CodeInput}
R> reco_mlr3 <- ctrml(base = bf_mat, hat = val_bf_mat, obs = val_hfbts, 
+                     agg_order = 24, agg_mat = matrix(1L, ncol = 7),
+                     approach = "mlr3", params = list(.key = "regr.xgboost"))
R> print(reco_mlr3, n_row = 2, n_col = 6)
\end{CodeInput}

\begin{CodeOutput}
      k-24 h-1   k-24 h-2  k-12 h-1   k-12 h-2   k-12 h-3   k-12 h-4
s-1 3450168.51 4002654.45 1560229.6 1889938.89 1857904.95 2144749.50
s-2   71943.42   76925.49   30351.8   41591.63   32877.66   44047.83
... (6 more rows, 114 more columns)
Use `print(reco_mlr3, n_row, n_col)` to see more rows and columns.
\end{CodeOutput}

\end{CodeChunk}

Model tuning can be achieved by the \code{param} and the \code{tuning}
argument. We illustrate parameter tuning using extreme gradient boosting
by tuning the maximum tree depth (\code{max_depth}) from \(2\) to \(6\)
(\code{paradox::to_tune(paradox::p_int(2, 6))}), and terminating
(\code{terminator}) the tuning process after \(10\) evaluations
(\code{mlr3tuning::trm("evals", n_evals = 10)}).

\begin{CodeChunk}

\begin{CodeInput}
R> md <- paradox::to_tune(paradox::p_int(2, 6))
R> trmp <- mlr3tuning::trm("evals", n_evals = 10)
R> reco_mlr3t <- ctrml(base = bf_mat, hat = val_bf_mat, obs = val_hfbts,
+                      agg_order = 24, agg_mat = matrix(1L, ncol = 7),
+                      approach = "mlr3", tuning = list(terminator = trmp),
+                      params = list(.key = "regr.xgboost", max_depth = md))
\end{CodeInput}

\end{CodeChunk}

We finally compare the performance of different methods on the test set
spanning June, 29 2025 to June, 30, 2025. Table~\ref{tbl-compare} shows
the RMSE ratios of reconciled forecasts of the series of the total of
Italy under the cross-temporal and temporal frameworks, using different
reconciliation methods, relative to the base forecast RMSE on the test
set.

\begin{table}

\centering{

\centering
\begin{tabular}[t]{r>{\raggedright\arraybackslash}p{1mm}rrrrr>{\raggedright\arraybackslash}p{1mm}rrrrr}
\toprule
\multicolumn{1}{c}{ } & \multicolumn{1}{c}{} & \multicolumn{5}{c}{One-day-ahead} & \multicolumn{1}{c}{} & \multicolumn{5}{c}{Two-day-ahead} \\
\cmidrule(l{3pt}r{3pt}){3-7} \cmidrule(l{3pt}r{3pt}){9-13}
\multicolumn{1}{c}{ } & \multicolumn{1}{c}{ } & \multicolumn{3}{c}{te} & \multicolumn{2}{c}{ct} & \multicolumn{1}{c}{ } & \multicolumn{3}{c}{te} & \multicolumn{2}{c}{ct} \\
\cmidrule(l{3pt}r{3pt}){3-5} \cmidrule(l{3pt}r{3pt}){6-7} \cmidrule(l{3pt}r{3pt}){9-11} \cmidrule(l{3pt}r{3pt}){12-13}
$k$ &  & ols & shr & wlsv & shr & ml &  & ols & shr & wlsv & shr & ml\\
\midrule
1 &  & 0.581 & 0.202 & 0.804 & 0.161 & 0.343 &  & 0.685 & 0.412 & 0.789 & 0.192 & 0.853\\
2 &  & 0.378 & 0.130 & 0.527 & 0.103 & 0.216 &  & 0.985 & 0.594 & 1.139 & 0.276 & 1.229\\
3 &  & 0.985 & 0.353 & 1.420 & 0.278 & 0.531 &  & 0.633 & 0.382 & 0.730 & 0.175 & 0.785\\
4 &  & 0.360 & 0.123 & 0.506 & 0.096 & 0.193 &  & 1.108 & 0.677 & 1.297 & 0.312 & 1.405\\
6 &  & 1.049 & 0.328 & 1.560 & 0.276 & 0.560 &  & 1.055 & 0.651 & 1.230 & 0.293 & 1.316\\
8 &  & 1.041 & 0.371 & 1.598 & 0.310 & 0.607 &  & 1.077 & 0.672 & 1.275 & 0.311 & 1.298\\
12 &  & 1.042 & 0.640 & 4.469 & 0.794 & 1.695 &  & 1.051 & 0.850 & 1.386 & 0.397 & 1.563\\
24 &  & 0.169 & 0.109 & 1.619 & 0.270 & 0.611 &  & 0.991 & 1.014 & 1.478 & 0.447 & 1.433\\
\bottomrule
\end{tabular}

}

\caption{\label{tbl-compare}Italian energy load: RMSE ratios of
reconciled forecasts for Italy under the cross-temporal (ct) and
temporal (te) frameworks, using different reconciliation methods,
relative to the base forecast RMSE on the test set. Reconciliation
methods include: ml (machine learning); ols (ordinary least squares);
shr (generalised least squares with shrunk covariance); and wlsv
(weighted least squares using series variances).}

\end{table}%

\section[Conclusion]{Conclusion} \label{sec:conclusion}
We present the \proglang{R} packages \pkg{FoReco} and \pkg{FoRecoML}, which perform   linear and non-linear ML-based forecast reconciliation, respectively.
Unlike existing forecast reconciliation packages, these two packages offer  
a unified toolbox that supports a wide range of reconciliation approaches across all three major reconciliation frameworks (cross-sectional, temporal and cross-temporal).
 All functions in \pkg{FoReco} and \pkg{FoRecoML} follow a consistent and user-friendly design principle -- "base forecasts in, reconciled forecasts out" -- making them particularly accessible to new users.
For expert users, all reconciliation functions are fully controllable, allowing detailed customization of the reconciliation methods to suit specific forecasting needs.

\section*{Acknowledgments}

We thank Tommaso Di Fonzo for providing helpful
comments on an earlier draft of this paper.
Ines Wilms and Yangzhuoran Fin Yang are supported by a grant from the Dutch Research Council (NWO; grant number \texttt{VI.Vidi.211.032}). Daniele Girolimetto is supported by the Ministero dell’Università e della Ricerca with the project PRIN2022 “PRICE: A New Paradigm for High Frequency Finance” (project number \texttt{2022C799SX}).

\bibliography{refs}

\end{document}